\documentclass[aps,prd,preprint,superscriptaddress,showpacs]{revtex4}

\usepackage{amsfonts}
\usepackage{mathrsfs}
\usepackage{amssymb}
\usepackage{amsmath}
\usepackage{graphicx,epsfig}

\def\bwt{\begin{widetext}}

\def\ewt{\end{widetext}}

\def\be{\begin{equation}}

\def\ee{\end{equation}}

\def\bea{\begin{eqnarray}}

\def\eea{\end{eqnarray}}

\def\bean{\begin{eqnarray*}}

\def\eean{\end{eqnarray*}}

\def\bary{\begin{array}}

\def\eary{\end{array}}

\def\bit{\begin{itemize}}

\def\eit{\end{itemize}}

\def\su5u1{SU(5) \times U(1)}

\def\fsu5u1{SU(5) \times U(1)'}

\def\so10{SO(10)}

\def\sq20{SO(10) \times SO(10)}

\def\bwt{\begin{widetext}}
\def\ewt{\end{widetext}}
\def\be{\begin{equation}}
\def\ee{\end{equation}}
\def\bea{\begin{eqnarray}}
\def\eea{\end{eqnarray}}
\def\bean{\begin{eqnarray*}}
\def\eean{\end{eqnarray*}}
\def\bary{\begin{array}}
\def\eary{\end{array}}
\def\bit{\begin{itemize}}
\def\eit{\end{itemize}}

\def\su5u1{SU(5) \times U(1)}
\def\fsu5u1{SU(5) \times U(1)'}
\def\so10{SO(10)}
\def\sq20{SO(10) \times SO(10)}

\usepackage{amssymb}

\begin{document}

\setlength{\parskip}{0cm}

\title{ General Gauge and Anomaly Mediated Supersymmetry Breaking
 in Grand Unified Theories with Vector-Like Particles }

\author{Tianjun Li}

\affiliation{Key Laboratory of Frontiers in Theoretical Physics,
      Institute of Theoretical Physics, Chinese Academy of Sciences,
Beijing 100190, P. R. China }

\affiliation{George P. and Cynthia W. Mitchell Institute for
Fundamental Physics and Astronomy, Texas A$\&$M University, College Station, TX
77843, USA }

\author{Dimitri V. Nanopoulos}

\affiliation{George P. and Cynthia W. Mitchell Institute for
Fundamental Physics and Astronomy, Texas A$\&$M University, College Station, TX
77843, USA }

\affiliation{Astroparticle Physics Group,
Houston Advanced Research Center (HARC),
Mitchell Campus, Woodlands, TX 77381, USA}

\affiliation{Academy of Athens, Division of Natural Sciences,
 28 Panepistimiou Avenue, Athens 10679, Greece }



\begin{abstract}

In Grand Unified Theories (GUTs) from orbifold and various 
string constructions the generic vector-like
particles do not need to form complete $SU(5)$ or $SO(10)$ 
representations. To realize them concretely, we present
 orbifold $ SU(5)$ models, orbifold $SO(10)$ models
 where the gauge symmetry can be broken down to flipped 
$SU(5)\times U(1)_X$ or Pati-Salam 
$SU(4)_C \times SU(2)_L \times SU(2)_R$ gauge symmetries, 
and F-theory $SU(5)$ models. Interestingly, these 
vector-like particles can be at the TeV-scale so
that the lightest CP-even Higgs boson mass can be lifted, or 
play the messenger fields in the Gauge Mediated Supersymmetry 
Breaking (GMSB). Considering GMSB, ultraviolet insensitive 
Anomaly Mediated Supersymmetry Breaking (AMSB), and 
the deflected AMSB, we study the general gaugino mass relations 
and their indices,  which are valid from the GUT scale to the 
electroweak scale at one loop, in the $SU(5)$ models, the 
flipped $SU(5)\times U(1)_X$ models, and the  Pati-Salam 
$SU(4)_C \times SU(2)_L \times SU(2)_R$ models. In the deflected 
AMSB, we also define the new indices for the gaugino 
mass relations, and calculate them as well.  
Using these gaugino mass relations and their indices, 
we may probe the messenger fields at intermediate
scale in the GMSB and deflected AMSB, determine 
the  supersymmetry breaking mediation mechanisms,
and distinguish the four-dimensional GUTs, orbifold GUTs, 
and F-theory GUTs.

\end{abstract}


\preprint{ACT-07-10, MIFPA-10-21}

\maketitle

\section{Introduction}

The supersymmetric Standard Model (SM) is the most elegant
extension of the SM since it solves the gauge hiearchy
problem naturally. In particular, the gauge coupling
unification can be achieved at about $2\times 10^{16}$~GeV~\cite{Ellis:1990zq}, 
and the lightest supersymmetric particle (LSP) like the neutralino can
be the cold dark matter candidate~\cite{Ellis:1983wd, Goldberg:1983nd}.
To solve the gauge hiearchy problem in the SM, supersymmetry should be broken around
the TeV scale.  Thus, at the Large Hadron Collider (LHC) and future International
Linear Collider (ILC), we may observe the  supersymmetric particles and 
get information about their mass spectra and interactions. 
The key questions are how to determine 
the mediation mechanisms for  supersymmetry breaking and how to probe the
Grand Unified Theories (GUTs) and string derived GUTs.

In the conventional supersymmetric SMs, supersymmetry is 
assumed to be broken in the hidden sector, and then its 
breaking effects are mediated to the SM observable sector. 
However, the relations between the supersymmetric particle 
spectra and the fundamental theories can be very complicated 
and model dependent. Interestingly, comparing to the  
supersymmetry breaking soft masses for squarks and sleptons, 
the gaugino masses have the simplest form and appear to be 
the least model dependent~\cite{Ellis:1985jn, Choi:2007ka}. 
For instance, with gravity mediated supersymmetry breaking
in GUTs, we have a universal gaugino mass $M_{1/2}$ 
at the GUT scale, which is called the minimal supergravity (mSUGRA)
scenario~\cite{mSUGRA}. Thus, we have the gauge coupling
relation and the gaugino mass relation at the GUT scale $M_{\rm GUT}$:
\begin{eqnarray}
{{1}\over {\alpha_3}} ~=~  {{1}\over {\alpha_2}} 
~=~  {{1}\over {\alpha_1}} ~,~\,
\label{mSUGRA-C}
\end{eqnarray}
\begin{eqnarray}
{{M_3}\over {\alpha_3}} ~=~  {{M_2}\over {\alpha_2}} 
~=~  {{M_1}\over {\alpha_1}} ~,~\,
\label{mSUGRA}
\end{eqnarray}
where $\alpha_3$, $\alpha_2$, and $\alpha_1\equiv 5\alpha_{Y}/3$ are gauge
couplings respectively for $SU(3)_C$, $SU(2)_L$, 
and $U(1)_Y$ gauge symmetries, and 
$M_3$, $M_2$, and $M_1$ are  the  masses 
 for $SU(3)_C$, $SU(2)_L$, and $U(1)_Y$ 
gauginos, respectively. Note that $M_i/\alpha_i$ are 
constant under one-loop renormalization group
equation (RGE) running, thus, we obtain that
the above  gaugino mass relation in Eq.~(\ref{mSUGRA}) is valid
 from the GUT scale to the electroweak scale at one loop.
Because the two-loop RGE running effects on gaugino masses
are very small, we can test this gaugino mass relation
at the LHC and ILC where the gaugino masses can be 
measured~\cite{Cho:2007qv, Barger:1999tn}.
Recently, considering the GUTs with high-dimensional 
operators~\cite{Ellis:1985jn, Hill:1983xh, Shafi:1983gz, Ellis:1984bm, Drees:1985bx, 
Anderson:1999uia, Chamoun:2001in, Chakrabortty:2008zk, Martin:2009ad, 
Bhattacharya:2009wv, Feldman:2009zc, Chamoun:2009nd}
and the F-theory GUTs with $U(1)$ fluxes~\cite{Vafa:1996xn,
Donagi:2008ca, Beasley:2008dc, Beasley:2008kw, Donagi:2008kj,
Font:2008id, Jiang:2009zza, Blumenhagen:2008aw, Jiang:2009za,
Li:2009cy, Leontaris:2009wi, Li:2010mr, Li:2010dp},
we generalized the mSUGRA scenario~\cite{Li:2010xr}. 
In particular, we studied
the generic gaugino mass relations and proposed 
their indices~\cite{Li:2010xr}. As we know,
there are three major supersymmetry breaking mediation
schemes: gravity medidated supersymmetry breaking~\cite{mSUGRA}, 
Gauge Mediated Supersymmetry Breaking 
(GMSB)~\cite{gaugemediation}, and
Anomaly Mediated Supersymmetry Breaking 
(AMSB)~\cite{anomalymediation, UVI-AMSB, D-AMSB}.
Thus, we shall study the generic
gaugino mass relations and their indices in
 the general GMSB and AMSB.

On the other hand, there exists a few pecent fine-tuning to have
the lightest CP-even Higgs boson mass heavier than 114 GeV in
the Minimal Supersymmetric Standard Model (MSSM). One possible
solution is that we introduce the TeV-scale vector-like 
particles~\cite{Moroi:1992zk}.
The lightest CP-even Higgs boson mass can be lifted due to
the large Yukawa couplings for these 
vector-like particles~\cite{Moroi:1992zk}.
Moreover, in the GMSB~\cite{gaugemediation} and deflected 
AMSB~\cite{D-AMSB}, we need messenger fields 
at the intermediate scale, which are also vector-like.  
Also, we can use the messenger fields to generate the
correct neutrino masses and mixings in the mean
 time~\cite{Joaquim:2006uz, Mohapatra:2008wx}. 
Thus, it is interesting to study the GUTs with generic vector-like 
particles. 

In this paper,  we first point out that the generic vector-like
particles do not need to form complete $SU(5)$ or $SO(10)$ 
representations  in GUTs from the orbifold 
constructions~\cite{kawa, GAFF, LHYN, AHJMR, Li:2001qs, Dermisek:2001hp,
Li:2001tx, Gogoladze:2003ci}, 
intersecting D-brane model building on Type II 
orientifolds~\cite{Blumenhagen:2005mu, Cvetic:2002pj, Chen:2006ip},
M-theory on $S^1/Z_2$ with Calabi-Yau 
compactifications~\cite{Braun:2005ux, Bouchard:2005ag}, and 
F-theory with $U(1)$ fluxes~\cite{Vafa:1996xn,
Donagi:2008ca, Beasley:2008dc, Beasley:2008kw, Donagi:2008kj,
Font:2008id, Jiang:2009zza, Blumenhagen:2008aw, Jiang:2009za,
Li:2009cy, Leontaris:2009wi, Li:2010mr, Li:2010dp}. 
Therefore, in the GMSB and deflected AMSB,
the messenger fields do not need to form complete $SU(5)$ or $SO(10)$ 
representations. The gauge coupling unification can be
preserved by introducing  the extra vector-like particles
at the intermediate scale that
do not mediate supersymmetry breaking. To be concrete, 
we present the 
 orbifold $ SU(5)$ models with additional vector-like particles, 
the  orbifold $SO(10)$ models with 
extra vector-like particles where the gauge symmetry
can be broken down to  flipped $SU(5)\times U(1)_X$ or
 Pati-Salam $SU(4)_C \times SU(2)_L \times SU(2)_R$
gauge symmetries, and the F-theory $SU(5)$ models
with generic vector-like particles. 
In short, these vector-like particles can be at the TeV scale 
so that we can increase the 
lightest CP-even Higgs boson mass in the MSSM~\cite{Moroi:1992zk}, 
and they can be
the messenger fields in the GMSB and deflected AMSB as well.
By the way, if the vector-like particles are around
the TeV scale, there may exist the possibility of flavour 
changing neutral currents even at tree level. To solve this
problem, we can require that the mixings between the
TeV-scale vector-like particles and the SM fermions are
very small.

In addition, we shall study the general gaugino mass relations
and their indices in the GMSB and AMSB, which are
valid from the GUT scale to the electroweak scale at one loop.
We briefly review the gaugino mass relations and their indices in
the generalization of the mSUGRA~\cite{Li:2010xr}, 
and define the suitable gaugino mass
relations in the GMSB and AMSB. For the GMSB, we first briefly
review the gaugino masses. With various possible
messenger fields, we calculate the gaugino
mass relations and their indices in the $SU(5)$ models, the 
flipped $SU(5)\times U(1)_X$ models, and the  Pati-Salam 
$SU(4)_C \times SU(2)_L \times SU(2)_R$ models. 
These kinds of models can be realized in 
orbifold GUTs, F-theory $SU(5)$ models with $U(1)_Y$ flux,
 F-theory $SO(10)$ models with $U(1)_X$ flux where the $SO(10)$
gauge symmetry is broken down to flipped $SU(5)\times U(1)_X$
gauge symmetries (we will denote them as 
F-theory flipped $SU(5)\times U(1)_X$
models), and F-theory $SO(10)$ models with
 $U(1)_{B-L}$ flux where the $SO(10)$ gauge symmetry is broken down
to $SU(3)_C\times SU(2)_L \times SU(2)_R \times U(1)_{B-L}$
gauge symmetries (we will denote them as  F-theory 
$SU(3)_C\times SU(2)_L \times SU(2)_R \times U(1)_{B-L}$ models). 
Using the gaugino mass relations and their indices,
we can probe the messenger fields at the imtermediate scale.
Moreover, for the AMSB, we first briefly review the
gaugino masses as well. To solve the tachyonic slepton problem for
the original AMSB, we consider two scenarios:
the ultraviolet (UV) insensitive AMSB~\cite{UVI-AMSB} and 
the deflected AMSB~\cite{D-AMSB}. 
In the UV insensitive AMSB, we calculate the gaugino mass
relations and their indices in the $SU(5)$ models with and
without the TeV-scale vector-like particles that form complete
$SU(5)$ multiplets, and in the flipped $SU(5)\times U(1)_X$
models with  TeV-scale vector-like particles that form complete
$SU(5)\times U(1)_X$ multiplets. To achieve the one-step gauge coupling
unification, we emphasize that the discussions for the
Pati-Salam models are similar to those in the $SU(5)$ models.
 In the deflected AMSB, without and with
the suitable TeV-scale vector-like particles that can lift
the lightest CP-even Higgs boson mass, we
study the generic gaugino mass relations and their indices
in the $SU(5)$ models, 
flipped $SU(5)\times U(1)_X$ models, and Pati-Salam 
$SU(4)_C \times SU(2)_L \times SU(2)_R$ models with various possible
messenger fields. To probe the messenger fields at 
intermediate scale, we define
the new indices for the gaugino mass relations, and
calculate them in details. Also, we find that in most of our scenarios,
the gluino can be the lightest gaugino at low energy.
In particular, we propose a new kind of interesting flipped
$SU(5)$ models as well.

 Furthermore, using the gaugino mass relations 
and their indices, we explain how to determine 
the  supersymmetry breaking mediation mechanisms,
 and how to probe the four-dimensional
GUTs, orbifold GUTs, and F-theory GUTs. 
Also, in order to distinguish between the different scenarios with the
same gaugino mass relations and the same indices,
we need to consider the squark and slepton
masses as well, which will be studied elsewhere~\cite{TLDN}.

This paper is organized as follows. In Sectin II,  we
discuss the vector-like particles that we are
interested in, and construct orbifold GUTs
and F-theory $SU(5)$ models with generic vector-like particles.
We briefly discuss the gaugino mass relations and their indices 
in Section III. We study the gaugino mass relations
and their indices for GMSB and AMSB in Section IV and
V, respectively. We consider the implications of
 the gaugino mass relations
and their indices in  Section VI.
Our conclusions are
given in Section VII. We briefly review the 
del Pezzo Surfaces in Appendix A.



\section{Generic Vector-Like Particles in the Orbifold and F-Theory GUTs}

In the GMSB and deflected AMSB, there exist messenger fields at
intermediate scales, which are vector-like particles. 
To realize 
gauge coupling unification, in the traditional GMSB and 
deflected AMSB, we assume that the messenger fields form
complete $SU(5)$ representations, for example,
$(\mathbf{5},~\mathbf{\overline{5}})$. However, we do not
have vector-like particles in 
complete $SU(5)$ representations in quite a few kinds of
 model building. In the intersecting D-brane 
model building on Type II orientifolds where  the $SU(5)$
gauge symmetry is broken down to the SM gauge symmetry by D-brane 
splitting~\cite{Blumenhagen:2005mu, Cvetic:2002pj, Chen:2006ip},
and in the M-theory on $S^1/Z_2$ with Calabi-Yau manifold
compactifications where the $SU(5)$ and $SO(10)$
 gauge symmetries are respectively
broken down to the $SU(3)_C\times SU(2)_L\times U(1)_Y$
 and $SU(3)_C\times SU(2)_L\times U(1)_{B-L}\times U(1)_{I3R}$
gauge symmetries by Wilson lines~\cite{Braun:2005ux, Bouchard:2005ag}, 
 we can not have the massless vector-like particles
that form complete GUT representations.
For the bulk vector-like particles in the orbifold 
GUTs~\cite{kawa, GAFF, LHYN, AHJMR, Li:2001qs, Dermisek:2001hp,
Li:2001tx, Gogoladze:2003ci},
we can not keep the zero modes for all the vector-like particles
in the complete GUT representations,
{\it i.e.}, the zero modes of some vector-like particles will
be projected out. In the F-theory 
GUTs~\cite{Vafa:1996xn, Donagi:2008ca, Beasley:2008dc, Beasley:2008kw, 
Donagi:2008kj, Font:2008id, Jiang:2009zza, Blumenhagen:2008aw, 
Jiang:2009za, Li:2009cy, Leontaris:2009wi, Li:2010mr, Li:2010dp}, 
we can also obtain the
 vector-like particles that do not form complete
GUT multiplets. In fact, the $SU(5)$ models,  
flipped $SU(5)\times U(1)_X$ models~\cite{smbarr, dimitri, AEHN-0, LNY,
Jiang:2006hf}, 
and $SU(3)_C\times SU(2)_L \times SU(2)_R \times U(1)_{B-L}$ models 
with additional vector-like particles have already been constructed
locally in F-theory~\cite{Beasley:2008dc, Beasley:2008kw, 
Font:2008id, Jiang:2009zza, Jiang:2009za, Li:2009cy, Li:2010dp}.
Interestingly, we should emphasize that
this is the reason why we can solve the
doublet-triplet splitting problem in these kinds of model building.
In this Section, we shall present the orbifold
 $ SU(5)$ models with additional vector-like particles, 
 the orbifold $SO(10)$ models with 
additional vector-like particles where the gauge symmetry
can be broken down to flipped $SU(5)\times U(1)_X$ or
 Pati-Salam $SU(4)_C \times SU(2)_L \times SU(2)_R$
gauge symmetries, and the F-theory $SU(5)$ models
with generic vector-like particles. 

First, let us explain our convention for supersymmetric SMs. 
We denote the left-handed quark doublets,  right-handed 
 up-type quarks, right-handed  down-type quarks, 
 left-handed lepton doublets, right-handed neutrinos
and right-handed charged leptons as $Q_i$,  $U^c_i$, $D^c_i$,
$L_i$, $N^c_i$, and  $E^c_i$, respectively. Also, we denote 
one pair of  Higgs doublets as $H_u$ and $H_d$, which give masses
to the up-type quarks/neutrinos and the down-type quark/charged
leptons, respectively.
In this paper, we consider the vector-like particles whose quantum 
numbers are the same as those of the SM fermions and their Hermitian 
conjugates, particles in the $SU(5)$ symmetric representation and 
their Hermitian conjugates, and the $SU(5)$ adjoint particles. 
Their quantum numbers under 
$SU(3)_C \times SU(2)_L \times U(1)_Y$ and their
contributions to one-loop beta functions
$\Delta b \equiv (\Delta b_1, \Delta b_2, \Delta b_3)$ as complete 
supermultiplets are given as follows
\begin{eqnarray}
&& XQ + XQ^c = {\mathbf{(3, 2, {1\over 6}) + ({\bar 3}, 2,
-{1\over 6})}}\,, \quad \Delta b =({1\over 5}, 3, 2)\,;\\ 
&& XU + XU^c = {\mathbf{ ({3},
1, {2\over 3}) + ({\bar 3},  1, -{2\over 3})}}\,, \quad \Delta b =
({8\over 5}, 0, 1)\,;\\ 
&& XD + XD^c = {\mathbf{ ({3},
1, -{1\over 3}) + ({\bar 3},  1, {1\over 3})}}\,, \quad \Delta b =
({2\over 5}, 0, 1)\,;\\  
&& XL + XL^c = {\mathbf{(1,  2, {1\over 2}) + ({1},  2,
-{1\over 2})}}\,, \quad \Delta b = ({3\over 5}, 1, 0)\,;\\ 
&& XE + XE^c = {\mathbf{({1},  1, {1}) + ({1},  1,
-{1})}}\,, \quad \Delta b = ({6\over 5}, 0, 0)\,;\\ 
&& XG = {\mathbf{({8}, 1, 0)}}\,, \quad \Delta b = (0, 0, 3)\,;\\ 
&& XW = {\mathbf{({1}, 3, 0)}}\,, \quad \Delta b = (0, 2, 0)\,;\\
&& XT + XT^c = {\mathbf{(1, 3, 1) + (1, 3,
-1)}}\,, \quad \Delta b =({{18}\over 5}, 4, 0)\,;\\ 
&& XS + XS^c = {\mathbf{(6,  1, -{2\over 3}) + ({\bar 6},
1, {2\over 3})}}\,, \quad \Delta b = ({16\over 5}, 0, 5)\,;\\ 
&& XY + XY^c = {\mathbf{(3, 2, -{5\over 6}) + ({\bar 3}, 2,
{5\over 6})}}\,, \quad \Delta b =(5, 3, 2)\,.\,
\end{eqnarray}

\subsection{Traditional Four-dimensional Grand Unified Theories}

First, let us briefly review the $SU(5)$ models and explain 
the convention. We define the $U(1)_{Y}$ hypercharge generator in $SU(5)$ as follows
\bea 
T_{\rm U(1)_{Y}}={\rm diag} \left(-{1\over 3}, -{1\over 3}, -{1\over 3},
 {1\over 2},  {1\over 2} \right)~.~\,
\label{u1y}
\eea
Under $SU(3)_C\times SU(2)_L \times U(1)_Y$ gauge symmetry, 
the $SU(5)$ representations
are decomposed as follows
\begin{eqnarray}
\mathbf{5} &=& \mathbf{(3, 1, -1/3)} \oplus \mathbf{(1, 2, 1/2)}~,~ \\
\mathbf{\overline{5}}  &=& 
\mathbf{(\overline{3}, 1, 1/3)} \oplus \mathbf{(1, 2, -1/2)}~,~ \\
\mathbf{10} &=& \mathbf{(3, 2, 1/6)} \oplus \mathbf{({\overline{3}}, 1, -2/3)}
\oplus \mathbf{(1, 1, 1)}~,~ \\
\mathbf{\overline{10}} &=& \mathbf{(\overline{3}, 2, -1/6)} 
\oplus \mathbf{(3, 1, 2/3)}
\oplus \mathbf{(1, 1, -1)}~,~ \\
\mathbf{24} &=&  \mathbf{(8, 1, 0)} \oplus  \mathbf{(1, 3, 0)} 
\oplus  \mathbf{(1, 1, 0)}  \oplus \mathbf{(3, 2, -5/6)} \oplus
 \mathbf{(\overline{3}, 2, 5/6)}~.~\,
\end{eqnarray}
There are three families of the SM fermions 
whose quantum numbers under $SU(5)$ are
\bea
F'_i=\mathbf{10},~ {\overline f}'_i={\mathbf{\bar 5}},~
N^c_i={\mathbf{1}}~,~
\label{SU(5)-smfermions}
\eea
where $i=1, 2, 3$ for three families. 
The SM particle assignments in $F'_i$ and ${\bar f}'_i$  are
\bea
F'_i=(Q_i, U^c_i, E^c_i)~,~{\overline f}'_i=(D^c_i, L_i)~.~
\label{SU(5)-smparticles}
\eea

To break the $SU(5)$ gauge symmetry and electroweak gauge symmetry, 
we introduce the adjoint Higgs field and one pair
of Higgs fields whose quantum numbers under $SU(5)$ are
\bea
\Phi'~=~ {\mathbf{24}}~,~~~
h'~=~{\mathbf{5}}~,~~~{\overline h}'~=~{\mathbf{\bar {5}}}~,~\,
\label{SU(5)-1-Higgse}
\eea
where $h'$ and ${\overline h}'$ contain the Higgs doublets
 $H_u$ and $H_d$, respectively.

Second, we would like to briefly review the flipped
$SU(5)\times U(1)_{X}$ models~\cite{smbarr, dimitri, AEHN-0}. 
The gauge group $SU(5)\times U(1)_{X}$ can be embedded 
into $SO(10)$.
We define the generator $U(1)_{Y'}$ in $SU(5)$ as 
\bea 
T_{\rm U(1)_{Y'}}={\rm diag} \left(-{1\over 3}, -{1\over 3}, -{1\over 3},
 {1\over 2},  {1\over 2} \right).
\label{u1yp}
\eea
The hypercharge is given by
\bea
Q_{Y} = {1\over 5} \left( Q_{X}-Q_{Y'} \right).
\label{ycharge}
\eea

There are three families of the SM fermions 
whose quantum numbers under $SU(5)\times U(1)_{X}$ are
\bea
F_i={\mathbf{(10, 1)}},~ {\bar f}_i={\mathbf{(\bar 5, -3)}},~
{\bar l}_i={\mathbf{(1, 5)}},
\label{smfermions}
\eea
where $i=1, 2, 3$. The particle assignments for the SM fermions are
\bea
F_i=(Q_i, D^c_i, N^c_i)~,~~{\overline f}_i=(U^c_i, L_i)~,~~{\overline l}_i=E^c_i~.~
\label{smparticles}
\eea

To break the GUT and electroweak gauge symmetries, we introduce two pairs
of Higgs fields whose quantum numbers under $SU(5)\times U(1)_X$ are
\bea
H={\mathbf{(10, 1)}}~,~~{\overline{H}}={\mathbf{({\overline{10}}, -1)}}~,~~
h={\mathbf{(5, -2)}}~,~~{\overline h}={\mathbf{({\bar {5}}, 2)}}~,~\,
\label{Higgse1}
\eea
where $h$ and ${\overline h}$ contain the Higgs doublets 
 $H_d$ and $H_u$, respectively.

Moreover, the flipped $SU(5)\times U(1)_X$ models can be embedded into
$SO(10)$. Under $SU(5)\times U(1)_X$ gauge symmetry, 
the $SO(10)$ representations are decomposed as follows
\begin{eqnarray}
\mathbf{10} &=& \mathbf{(5, -2)} \oplus 
\mathbf{(\overline{5}, 2)}  ~,~ \\
\mathbf{{16}} &=&  \mathbf{(10, 1)} 
\oplus  \mathbf{(\overline{5}, -3)} \oplus 
\mathbf{(1, 5)} ~,~\\
\mathbf{45} &=&  \mathbf{(24, 0)} \oplus  \mathbf{ (1, 0)} 
\oplus  \mathbf{(10, -4)}  \oplus  \mathbf{(\overline{10}, 4)} 
~.~\,
\end{eqnarray}
Let us consider the
 vector-like particles which form complete 
flipped $SU(5)\times U(1)_X$ multiplets.
The quantum numbers for these additional vector-like particles
 under the $SU(5)\times U(1)_X$ gauge symmetry are
\begin{eqnarray}
&& XF ~=~{\mathbf{(10, 1)}}~,~~{\overline{XF}}~=~{\mathbf{({\overline{10}}, -1)}}~,~\\
&& Xf~=~{\mathbf{(5, 3)}}~,~~{\overline{Xf}}~=~{\mathbf{({\overline{5}}, -3)}}~,~\\
&& Xl~=~{\mathbf{(1, -5)}}~,~~{\overline{Xl}}~=~{\mathbf{(1, 5)}}~,~\\
&& Xh~={\mathbf{(5, -2)}}~,~~{\overline{Xh}}~=~{\mathbf{({\overline{5}}, 2)}} \\
&& XGW~=~ \mathbf{(24, 0)} ~,~~ XN~=~\mathbf{ (1, 0)} ~,~ \\
&& XX~=~\mathbf{(10, -4)}~,~~\overline{XX}~=~ \mathbf{(\overline{10}, 4)} ~.~\,
\end{eqnarray}

Moreover,  the particle contents for the decompositions
of $XF$, ${\overline{XF}}$, $Xf$, ${\overline{Xf}}$,
$Xl$, ${\overline{Xl}}$, $Xh$,  ${\overline{Xh}}$,
$XGW$, $XX$, and $\overline{XX}$ under the SM
gauge symmetries are
\begin{eqnarray}
&& XF ~=~ (XQ, XD^c, XN^c)~,~~ {\overline{XF}}~=~(XQ^c, XD, XN)~,~\\
&& Xf~=~(XU, XL^c)~,~ {\overline{Xf}}~=~ (XU^c, XL)~,~\\
&& Xl~=~ XE~,~ {\overline{Xl}}~=~ XE^c~,~\\
&& Xh~=~(XD, XL)~,~ {\overline{Xh}}~=~ (XD^c, XL^c)~,~\\
&& XGW~=~(XG, XW, XQ, XQ^c)~,~~ \\
&& XX ~=~ (XY, XU^c, XE)~,~~ {\overline{XX}}~=~(XY^c, XU, XE^c)~.~\,
\end{eqnarray}

In flipped $SU(5)\times U(1)_X$ models of
$SO(10)$ origin, there are two steps for gauge coupling unification:
the $SU(3)_C \times SU(2)_L$ gauge symmeties are unified first
at the scale $M_{32}$, and then the
 $SU(5)\times U(1)_X$ gauge symmetries
are unified at the higher scale $M_U$,
where $M_{32}$ is about the usual GUT scale  around
$2\times 10^{16}$ GeV. Thus,  the
condition for gauge coupling unification in the flipped
$SU(5)\times U(1)_X$ models can be relaxed elegantly.
To realize the string-scale gauge coupling unification
in the free fermionic string constructions~\cite{LNY} 
or the decoupling scenario in the F-theory model 
building~\cite{Jiang:2009zza, Jiang:2009za},
we introduce the TeV-scale vector-like particles which
 form the complete flipped $SU(5)\times U(1)_X$ multiplets~\cite{Jiang:2006hf}.
To avoid the Landau pole problem for the strong coupling,
we show that at the TeV scale, we can only introduce
the vector-like particles $(XF,~{\overline{XF}})$ or
$(XF,~{\overline{XF}})\oplus (Xl,~{\overline{Xl}})$~\cite{Jiang:2006hf}.
The flipped $SU(5)\times U(1)_X$ models with 
these vector-like particles are dubbed as the 
testable flipped $SU(5)\times U(1)_X$ models since they can
solve the monopole problem, realize the hybrid inflation,
lift the lightest CP-even Higgs boson mass, and
predict the proton decay within the reach of the
future proton decay experiments, etc~\cite{Jiang:2009za, Jiang:2006hf}.

Third, we would like to briefly review the 
Pati-Salam models. The gauge group  is
$SU(4)_C \times SU(2)_L \times SU(2)_R$, which can also be embedded 
into $SO(10)$.
There are three families of the SM fermions 
whose quantum numbers under $SU(4)_C \times SU(2)_L \times SU(2)_R$ are
\bea
F^L_i={\mathbf{(4, 2, 1)}}~,~~ F^{Rc}_i={\mathbf{(\overline{4}, 1, 2)}}~,~\,
\eea
where $i=1, 2, 3$.
Also, the particle assignments for the SM fermions are
\bea
F^L_i=(Q_i, L_i)~,~~F^{Rc}_i=(U^c_i, D^c_i, E^c_i, N^c_i)~.~\,
\eea

To break the Pati-Salam and electroweak gauge symmetries, 
we introduce one pair
of Higgs fields and one bidoublet Higgs field whose
quantum numbers under $SU(4)_C \times SU(2)_L \times SU(2)_R$ are
\bea
\Phi={\mathbf{(4, 1, 2)}}~,~~{\overline{\Phi}}={\mathbf{({\overline{4}}, 1, 2)}}~,~~
H'={\mathbf{(1, 2, 2)}}~,~\,
\label{Higgse1}
\eea
where $H'$ contains one pair of the Higgs doublets 
 $H_d$ and $H_u$.

Moreover, the Pati-Salam models can be embedded into
$SO(10)$ models. Under $SU(4)_C\times SU(2)_L\times SU(2)_R$ gauge symmetry, 
the $SO(10)$ representations are decomposed as follows
\begin{eqnarray}
\mathbf{10} &=& \mathbf{(6, 1, 1)} \oplus 
\mathbf{(1, 2, 2)}  ~,~ \\
\mathbf{{16}} &=&  \mathbf{(4, 2, 1)} 
\oplus  \mathbf{(\overline{4}, 1, 2)}  ~,~\\
\mathbf{45} &=&  \mathbf{(15, 1, 1)} \oplus  \mathbf{ (1, 3, 1)}  
\oplus  \mathbf{ (1, 1, 3)}
\oplus  \mathbf{(6, 2, 2)}  ~.~\,
\end{eqnarray}
Let us consider the
 vector-like particles which form complete 
 $SU(4)_C\times SU(2)_L \times SU(2)_R$ representations.
The quantum numbers for the  vector-like particles
 under the $SU(4)_C\times SU(2)_L\times SU(2)_R$ gauge symmetry are
\begin{eqnarray}
&& XFL ~=~{\mathbf{(4, 2, 1)}}~,~~
{\overline{XFL}}~=~{\mathbf{({\overline{4}}, 2, 1)}}~,~\\
&& XFR~=~{\mathbf{(4, 1, 2)}}~,~~
{\overline{XFR}}~=~{\mathbf{({\overline{4}}, 1, 2)}}~,~\\
&&
XD\overline{D}~=~\mathbf{(6, 1, 1)}~,~~XL\overline{L}~=~\mathbf{(1, 2, 2)}~,~\\
&& XG4~=~  \mathbf{(15, 1, 1)}~,~~ XWL~=~ \mathbf{ (1, 3, 1)}~,~\\
&& XWR~=~ \mathbf{ (1, 1, 3)}~,~~XZ={\mathbf{(6, 2, 2)}}~.~\,
\end{eqnarray}
Also,  the particle contents for the decompositions of
 $XFL$, ${\overline{XFL}}$,
 $XFR$, ${\overline{XFR}}$, $XD\overline{D}$, $XL\overline{L}$,
 $XG4$, $XWL$, $XWR$ and $XZ$ 
under the SM gauge symmetries are
\begin{eqnarray}
&& XFL ~=~ (XQ, XL)~,~~ {\overline{XFL}}~=~(XQ^c, XL^c)~,~\\
&& XFR~=~(XU, XD, XE, XN)~,~~ {\overline{XFR}}~=~ (XU^c, XD^c, XE^c, XN^c)~,~\\
&& XD\overline{D}~=~(XD, XD^c)~,~~XL\overline{L}~=~(XL, XL^c)~,~\\
&& XG4~=~(XG, XU, XU^c)~,~~XWL~=~XW \\
&& XWR~=~(XE, XE^c, XN) ~,~~XZ=(XQ, XQ^c, XY, XY^c)~.~
\end{eqnarray}



\subsection{Obifold Grand Unified Theories with Generic
Vector-Like Particles}

In the five-dimensional orbifold supersymmetric 
GUTs~\cite{kawa, GAFF, LHYN, AHJMR, Li:2001qs, Dermisek:2001hp,
Li:2001tx, Gogoladze:2003ci}, the five-dimensional manifold is
factorized into the product of ordinary four-dimensional Minkowski
space-time $M^4$ and the orbifold $S^1/(Z_2\times Z'_2)$.  The
corresponding coordinates are $x^{\mu}$ ($\mu = 0, 1, 2, 3$) and
$y\equiv x^5$. The radius for the fifth dimension is $R$.  The
orbifold $S^1/(Z_2\times Z'_2)$ is obtained by $S^1$ moduloing the
equivalent class
\begin{eqnarray}
P:~~y \sim -y~,~P':~~ y' \sim -y'~,~\,
\end{eqnarray}
where $y'\equiv y-\pi R/2$.  There are two fixed points, $y=0$ and
$y=\pi R/2$.

The $N=1$ supersymmetric theory in five dimensions have 8 real
supercharges, corresponding to $N=2$ supersymmetry in four dimensions.  In
terms of the physical degrees of freedom, the vector multiplet
contains a vector boson $A_M$ with $M=0, 1, 2, 3, 5$, two Weyl
gauginos $\lambda_{1,2}$, and a real scalar $\sigma$.  In the
four-dimensional $N=1$ supersymmetry language, it contains a vector
multiplet $V \equiv (A_{\mu}, \lambda_1)$ and a chiral multiplet
$\Sigma \equiv ((\sigma+iA_5)/\sqrt 2, \lambda_2)$ which transform in
the adjoint representation of group $G$.  The five-dimensional
hypermultiplet consists of two complex scalars $\phi$ and $\phi^c$,
and a Dirac fermion $\Psi$.  It can be decomposed into two chiral
mupltiplets $\Phi(\phi, \psi \equiv \Psi_R)$ and $\Phi^c(\phi^c,
\psi^c \equiv \Psi_L)$, which are in the conjugate representations of
each other under the gauge group.

The general action for the group $G$ gauge fields and their
couplings to the bulk hypermultiplet $\Phi$ is~\cite{NAHGW}
\begin{eqnarray}
S&=&\int{d^5x}\frac{1}{k g^2}
{\rm Tr}\left[\frac{1}{4}\int{d^2\theta} \left(W^\alpha W_\alpha+{\rm H.
C.}\right)
\right.\nonumber\\&&\left.
+\int{d^4\theta}\left((\sqrt{2}\partial_5+ {\bar \Sigma })
e^{-V}(-\sqrt{2}\partial_5+\Sigma )e^V+
\partial_5 e^{-V}\partial_5 e^V\right)\right]
\nonumber\\&&
+\int{d^5x} \left[ \int{d^4\theta} \left( {\Phi}^c e^V {\bar \Phi}^c +
 {\bar \Phi} e^{-V} \Phi \right)
\right.\nonumber\\&&\left.
+ \int{d^2\theta} \left( {\Phi}^c (\partial_5 -{1\over {\sqrt 2}} \Sigma)
\Phi + {\rm H. C.}
\right)\right]~.~\,
\end{eqnarray}

Under the parity operator $P$, the vector multiplet transforms as
\begin{eqnarray}
V(x^{\mu},y)&\to  V(x^{\mu},-y) = P V(x^{\mu}, y) P^{-1}
~,~\,
\end{eqnarray}
\begin{eqnarray}
 \Sigma(x^{\mu},y) &\to\Sigma(x^{\mu},-y) = - P \Sigma(x^{\mu}, y) P^{-1}
~.~\,
\end{eqnarray}
For the hypermultiplet $\Phi$ and $\Phi^c$, we have
\begin{eqnarray}
\Phi(x^{\mu},y)&\to \Phi(x^{\mu}, -y)  = \eta_{\Phi} P^{l_\Phi} \Phi(x^{\mu},y)
(P^{-1})^{m_\Phi}~,~\,
\end{eqnarray}
\begin{eqnarray}
\Phi^c(x^{\mu},y) &\to \Phi^c(x^{\mu}, -y)  = -\eta_{\Phi} P^{l_\Phi}
\Phi^c(x^{\mu},y) (P^{-1})^{m_\Phi}
~,~\,
\end{eqnarray}
where $\eta_{\Phi}$ is $\pm$, $l_{\Phi}$ and $m_{\Phi}$ are
respectively the numbers of the fundamental index and anti-fundamental
index for the bulk multiplet $\Phi$ under the bulk gauge group $G$.
For example, if $G$ is an $SU(N)$ group, for a fundamental
representation, we have $l_{\Phi}=1$ and $m_{\Phi}=0$, and for an adjoint
representation, we have $l_{\Phi}=1$ and $m_{\Phi}=1$.  Moreover, the
transformation properties for the vector multiplet and hypermultiplets
under $P'$ are the same as those under $P$.

For $G=SU(5)$, to break the $SU(5)$ gauge symmetry, we choose the
following $5\times 5$ matrix representations for the parity operators
$P$ and $P'$
\begin{equation}
\label{eq:pppsu5}
P={\rm diag}(+1, +1, +1, +1, +1)~,~P'={\rm diag}(+1, +1, +1, -1, -1)
 ~.~\,
\end{equation}
Under the $P'$ parity, the gauge generators $T^\alpha$ ($\alpha =1$,
$2$, ..., $24$) for $SU(5)$ are separated into two sets: $T^a$ are the
generators for the SM gauge group, and $T^{\hat a}$ are the generators
for the broken gauge group
\begin{equation}
P~T^a~P^{-1}= T^a ~,~ P~T^{\hat a}~P^{-1}= T^{\hat a}
~,~\,
\end{equation}
\begin{equation}
P'~T^a~P^{'-1}= T^a ~,~ P'~T^{\hat a}~P^{'-1}= - T^{\hat a}
~.~\,
\end{equation}
The zero modes of the $SU(5)/SM$ gauge bosons are projected out, thus,
the five-dimensional $N=1$ supersymmetric $SU(5)$ gauge symmetry is
broken down to the four-dimensional $N=1$ supersymmetric SM gauge
symmetry for the zero modes. For the zero modes and KK modes, the
four-dimensional $N=1$ supersymmetry is preserved on the 3-branes at both
fixed points, and only the SM gauge symmetry is preserved on the
3-brane at $y=\pi R/2$~\cite{Li:2001tx}.

For $G=SO(10)$, the generators $T^\alpha$ of $SO(10)$ are imaginary
antisymmetric $10 \times 10$ matrices.  In terms of the $2\times 2$
identity matrix $\sigma_0$ and the Pauli matrices $\sigma_i$, they can
be written as tensor products of $2\times 2$ and $5 \times 5$
matrices, $(\sigma_0, \sigma_1, \sigma_3) \otimes A_5$ and $\sigma_2
\otimes S_5$ as a complete set, where $A_5$ and $S_5$ are the $5\times
5$ real anti-symmetric and symmetric matrices.  The generators of the
$\su5u1$ gauge symmetries are
\bea
&& \sigma_0 \otimes A_3\,, \quad \sigma_0 \otimes A_2\,, \quad \sigma_0
\otimes A_X \nonumber \\
&& \sigma_2 \otimes S_3\,, \quad \sigma_2 \otimes S_2\,, \quad
\sigma_2 \otimes S_X \,,
\eea
the generators for flipped $SU(5)\times U(1)_X$ gauge symmetries are
\bea
&& \sigma_0 \otimes A_3\,, \quad \sigma_0 \otimes A_2\,, \quad
\sigma_1 \otimes A_X \nonumber \\
&& \sigma_2 \otimes S_3\,, \quad \sigma_2 \otimes S_2\,, \quad
\sigma_3 \otimes A_X \,,
\eea
and  the generators for Pati-Salam 
 $SU(4)_C \times SU(2)_L \times SU(2)_R$ gauge symmetries are
\begin{equation}
\begin{array}{cc}
  (\sigma_0,\sigma_1,\sigma_3) \otimes A_3~,~ &
     (\sigma_0,\sigma_1,\sigma_3) \otimes A_2~,~  \\
  \sigma_2 \otimes S_3~,~ & \sigma_2 \otimes S_2~,~\,
\end{array}
\label{eq:422gen}
\end{equation}
where $A_3$ and $S_3$ are respectively the diagonal blocks of $A_5$
and $S_5$ that have indices 1, 2, and 3, while the diagonal blocks
$A_2$ and $S_2$ have indices 4 and 5. $A_X$ and $S_X$ are the 
off-diagonal blocks of $A_5$ and $S_5$.

We choose the $10\times 10 $ matrix  for $P$ as
\begin{eqnarray}
P&=& \sigma_0 \otimes {\rm diag}(1,1,1,1,1)~.~\,
\end{eqnarray}
To break the $SO(10)$ down to $SU(5)\times U(1)$, we choose
\begin{eqnarray}
\label{eq:psu5}
P'&=& \sigma_2 \otimes {\rm diag}(1,1,1,1,1)~,~\,
\end{eqnarray}
to break the $SO(10)$ down to flipped $SU(5)\times U(1)_X$, we choose
\begin{eqnarray}
\label{eq:pfsu5}
P'&=& \sigma_2 \otimes {\rm diag}(1,1,1,-1,-1)~,~\,
\end{eqnarray}
and to break the $SO(10)$ down to the Pati-Salam gauge symmetries, we choose
\begin{eqnarray}
\label{eq:pps}
P'&=& \sigma_0 \otimes {\rm diag}(1,1,1,-1,-1)~.~\,
\end{eqnarray}
For the zero modes, the five-dimensional $N=1$ supersymmetric $SO(10)$
gauge symmetry is broken down to the four-dimensional $N=1$
supersymmetric $SU(5)\times U(1)$, flipped $SU(5)\times U(1)_X$ and Pati-Salam
$SU(4)_C\times SU(2)_L\times SU(2)_R$ gauge
symmetries.  Including the KK modes, the 3-branes at both fixed points
preserve the four-dimensional $N=1$ supersymmetry, and the gauge symmetry
on the 3-brane at $y=\pi R/2$ is $SU(5)\times U(1)$, flipped $SU(5)\times U(1)_X$
and Pati-Salam gauge symmetries, for different choices of $P'$~\cite{Li:2001tx}.


\begin{table}[htb]
\begin{center}
\begin{tabular}{|c|c|c||c|c|c|}
\hline
${\rm Representation }$ & $\eta_{\Phi}$ & ${\rm Zero~Modes}$ & 
${\rm Representation }$ & $\eta_{\Phi}$ & ~${\rm Zero~Modes}$ ~
\\\hline
$({\mathbf{5}}, ~{\mathbf{\overline{5}}})$ & $+1$ 
& $(XD, ~XD^c)$ & $({\mathbf{5}}, ~{\mathbf{\overline{5}}})$ &
$-1$ & $(XL, ~XL^c)$\\\hline
$({\mathbf{10}}, ~{\mathbf{\overline{10}}})$ & $+1$ 
& $(XU, ~XU^c),~ (XE, ~XE^c)$ & $({\mathbf{10}}, ~{\mathbf{\overline{10}}})$ &
$-1$ & $(XQ, ~XQ^c)$\\\hline
$({\mathbf{15}}, ~{\mathbf{\overline{15}}})$ & $+1$ 
& $(XT, ~XT^c),~ (XS, ~XS^c)$ & $({\mathbf{15}}, ~{\mathbf{\overline{15}}})$ &
$-1$ & $(XQ, ~XQ^c)$\\\hline
${\mathbf{24}}$ & $+1$ 
& $XG,~XW $ & ${\mathbf{24}} $ &
$-1$ & $(XY, ~XY^c)$\\\hline
\end{tabular}
\end{center}
\caption{ The possible vector-like particles which remain as zero modes
after orbifold projections in the orbifold $SU(5)$ models.}
\label{SUV-VP}
\end{table}


In Table~\ref{SUV-VP}, Table~\ref{SUT-F}, and Table~\ref{SUT-PS}, 
we present the possible vector-like particles, which remain as zero modes
after orbifold projections, in the orbifold $SU(5)$ models, in the orbifold
$SO(10)$ models whose gauge symmetry is broken down to the 
flipped $SU(5)\times U(1)_X$  gauge symmetry 
by orbifold projections, and the orbifold
$SO(10)$ models whose gauge symmetry is broken down to the
Pati-Salam $SU(4)_C\times SU(2)_L \times SU(2)_R $  gauge symmetry 
by orbifold projections,
respectively.


\begin{table}[htb]
\begin{center}
\begin{tabular}{|c|c|c||c|c|c|}
\hline
${\rm Representation }$ & $\eta_{\Phi}$ & ~${\rm Zero~Modes}$~ & 
${\rm Representation }$ & $\eta_{\Phi}$ & ${\rm Zero~Modes}$ 
\\\hline
${\mathbf{10}}$ & $+1$ 
& $Xh$ & ${\mathbf{10}}$ &
$-1$ & $\overline{Xh}$\\\hline
$({\mathbf{16}}, ~{\mathbf{\overline{16}}})$ & $+1$ 
& $(XF, ~\overline{XF})$ & $({\mathbf{16}}, ~{\mathbf{\overline{16}}})$ &
$-1$ & $(Xf, ~\overline{Xf}),~(Xl, ~\overline{Xl})$\\\hline
${\mathbf{45}}$ & $+1$ 
& $XGW,~XN $ & ${\mathbf{45}} $ &
$-1$ & $(XX, ~\overline{XX})$\\\hline
\end{tabular}
\end{center}
\caption{ The possible vector-like particles
which remain as zero modes after orbifold projections
 in the orbifold $SO(10)$ models
where the gauge symmetry is broken down to the flipped
$SU(5)\times U(1)_X $ gauge symmetries.}
\label{SUT-F}
\end{table}



\begin{table}[htb]
\begin{center}
\begin{tabular}{|c|c|c||c|c|c|}
\hline
${\rm Representation }$ & $\eta_{\Phi}$ & ${\rm Zero~Modes}$ & 
${\rm Representation }$ & $\eta_{\Phi}$ & ${\rm Zero~Modes}$ 
\\\hline
${\mathbf{10}}$ & $+1$ 
& $XD\overline{D}$ & ${\mathbf{10}}$ &
$-1$ & $XL\overline{L}$\\\hline
$({\mathbf{16}}, ~{\mathbf{\overline{16}}})$ & $+1$ 
& $(XFL, ~\overline{XFL})$ & $({\mathbf{16}}, ~{\mathbf{\overline{16}}})$ &
$-1$ & $(XFR, ~\overline{XFR})$\\\hline
${\mathbf{45}}$ & $+1$ 
& $XG4,~XWL,~XWR $ & ${\mathbf{45}} $ &
$-1$ & $XZ$\\\hline
\end{tabular}
\end{center}
\caption{ The possible vector-like particles 
which remain as zero modes after orbifold projections
 in the orbifold $SO(10)$ models
where the gauge symmetry is broken down to the Pati-Salam 
$SU(4)_C \times SU(2)_L \times SU(2)_R $ gauge symmetries.}
\label{SUT-PS}
\end{table}



\subsection{F-Theory $SU(5)$ Models with Generic Vector-Like Particles}

We first briefly review the F-theory model 
building~\cite{Vafa:1996xn, Donagi:2008ca,
Beasley:2008dc, Beasley:2008kw, Donagi:2008kj}.
The twelve-dimensional F theory is a convenient way to describe 
Type IIB vacua with varying
axion-dilaton $\tau=a+ie^{-\phi}$. We compactify F-theory on a
Calabi-Yau fourfold, which is elliptically fibered $\pi: Y_4 \to B_3$
with a section $\sigma: B_3 \to Y_4$. The base $B_3$ is the internal
space dimensions in Type IIB string theory, and the complex structure
of the $T^2$ fibre encodes $\tau$ at each point of $B_3$. The SM or GUT
gauge theories are on the worldvolume of the observable
seven-branes that wrap
a complex codimension-one suface in $B_3$. Denoting the complex
coordinate transverse to these seven-branes in $B_3$ as $z$, we can
write the elliptic fibration in  Weierstrass form
\begin{eqnarray}
 y^2=x^3+f(z)x+g(z)~,~\,
\end{eqnarray}
where $f(z)$  and $g(z)$ are sections of $K_{B_3}^{-4}$ and
$K_{B_3}^{-6}$, respectively. The complex structure of the fibre is 
\begin{eqnarray}
 j(\tau)~=~ {{4(24f)^3}\over {\Delta}}~,~~~
\Delta~=~ 4 f^3 + 27 g^2 ~.~\,
\end{eqnarray}
At the discriminant locus $\{\Delta=0\} \subset B_3$,
the torus $T^2$ degenerates by pinching one of its
cycles and becomes singular. For a generic pinching one-cycle 
 $(p, q)=p\alpha+q\beta$ where $\alpha$ and $\beta$
are one-cylces for the torus $T^2$, we obtain a $(p,q)$ seven-brane in 
the locus where the $(p,q)$ string can end.
The singularity types of the ellitically fibres fall into the 
familiar $ADE$ classifications, and we identify the corresponding 
$ADE$ gauge groups on the seven-brane world-volume. 
This is one of the most important advantages for the F-theory model building: 
the exceptional gauge groups appear rather naturally, which is absent in
perturbative Type II string theory. And then all the SM fermion Yuakwa
couplings in the GUTs can be generated.

We assume that the observable seven-branes with GUTs
on its worldvolume wrap a complex codimension-one 
suface $S$ in $B_3$, and the observable gauge symmetry
is $G_S$. When $h^{1,0}(S)\not=0$, the low energy
spectrum may contain the extra states obtained
by reduction of the bulk supergravity modes of
compactification. So we require that 
$\pi_1(S)$ be a finite group. In order to decouple
gravity and construct models locally, the extension 
of the local metric on $S$ to
a local Calabi-Yau fourfold must have a limit where
the surface $S$ can be shrunk to zero size. This implies
that the anti-canonical bundle on $S$ must be ample. 
Therefore, $S$ is a del Pezzo $n$ surface $dP_n$ with $n \ge 2$
in which $h^{2,0}(S)=0$ (for a brief review of del Pezzo surfaces,
see Appendix A).
By the way, the Hirzebruch surfaces with degree larger than 2
satisfy $h^{2,0}(S)=0$ but do not define the fully 
consistent decoupled models~\cite{Beasley:2008dc, Beasley:2008kw}.

 To describe the spectrum, we have to study the gauge theory 
of the worldvolume on the seven-branes.  We 
start from the maximal supersymmetric gauge theory on
$\mathbb{R}^{3,1}\times \mathbb{C}^{2}$ and then replace
$\mathbb{C}^{2}$ with the K\"ahler surface $S$. In order to have
four-dimensional ${\cal N}=1$ supersymmetry, the
maximal supersymmetric gauge theory on $\mathbb{R}^{3,1}\times
\mathbb{C}^{2}$ should be twisted. It was shown that there exists a
unique twist preserving ${\cal N}=1$ supersymmetry in four
dimensions, and chiral matters can arise from the bulk $S$ or the
codimension-one curve $\Sigma$ in $S$ which is the intersection
between the observable seven-branes and 
the other seven-brane(s)~\cite{Beasley:2008dc, Beasley:2008kw}.

In order to have the matter fields on $S$,
we consider a non-trivial vector bundle on $S$ with
a structure group $H_S$ which is a subgroup of $G_S$. Then the gauge
group $G_S$ is broken down to $\Gamma_S\times H_S$, and the adjoint
representation ${\rm ad}(G_S)$ of the $G_S$ is decomposed as 
\begin{equation}
{\rm ad}(G_S)\rightarrow
{\rm ad}(\Gamma_S)\bigoplus {\rm ad}(H_S)\bigoplus_j(\tau_j,T_j)~.~\,
\end{equation}
Employing the vanishing theorem of the del Pezzo surfaces,
we obtain the numbers of the generations and anti-generations 
by calculating the zero modes of the Dirac operator on $S$
\begin{eqnarray}
 n_{\tau_j}~=~ -\chi (S, \mathbf{T_j})~,~~~ 
n_{\tau_j^*}~=~ -\chi (S, \mathbf{T_j}^*)~,~\,
\end{eqnarray}
where $\mathbf{T_j}$ is the vector bundle on $S$ whose 
sections transform in the representation $T_j$ of $H_S$,
and $\mathbf{T_j}^*$ is the dual bundle of $\mathbf{T_j}$.
In particular, when the $H_S$ bundle is a line bundle $L$,
we have
\begin{eqnarray}
n_{\tau_j}~=~-\chi (S, L^j)~=~
-\Big[1+\frac{1}{2}\big(\int_{S}c_{1}({L}^{j})c_{1}(S)+
\int_{S}c_{1}({L}^{j})^2\big)\Big]~.~\,
\label{EulerChar}
\end{eqnarray}
In order to preserve supersymmetry, the line bundle $L$ should satisfy
 the BPS equation~\cite{Beasley:2008dc}
\begin{equation}
J_{S}\wedge c_{1}(L)=0,\label{BPS}
\end{equation}
where $J_{S}$ is the K\"ahler form on $S$. Moreover,
the admissible supersymmetric line bundles on del Pezzo surfaces must
satisfy $c_{1}(L)c_{1}(S)=0$, thus,
 $n_{\tau_j}=n_{\tau_j^*}$ and only the vector-like particles
can be obtained. In short, we can not have the chiral matter fields
on the worldvolume of the observable seven-branes.

Interestingly, the chiral superfields can come from the intersections
between the observable seven-branes and the other 
seven-brane(s)~\cite{Beasley:2008dc, Beasley:2008kw}. 
Let us consider a stack of seven-branes with gauge group
$G_{S'}$ that wrap a codimension-one surface $S'$ in $B_3$.
The intersection of $S$ and $S'$ is a codimenion-one curve
(Riemann surface) $\Sigma$ in $S$ and $S'$, 
and the gauge symmetry on $\Sigma$
will be enhanced to $G_{\Sigma}$ 
where $G_{\Sigma}\supset G_{S}\times G_{S'}$.
On this curve, there exist chiral matters from 
the decomposition of the adjoint representation
${\rm ad}G_{\Sigma}$ of $G_{\Sigma}$ as follows
\begin{equation}
{\rm ad}G_{\Sigma}={\rm ad}G_{S}\oplus {\rm ad}G_{S'}\oplus_{k}
({ U}_{k}\otimes { U'}_{k})~.~\,
\end{equation}
Turning on the non-trivial gauge bundles on $S$ and $S'$ respectively
with structure groups $H_S$ and $H_{S'}$, we break the gauge 
group $G_S\times G_{S'}$ down to the commutant subgroup 
$\Gamma_{S}\times\Gamma_{S'}$. Defining 
$\Gamma \equiv \Gamma_{S}\times\Gamma_{S'}$ and 
$H \equiv H_{S}\times H_{S'}$,
we can decompose ${ U}\otimes { U'}$ into the irreducible
representations as follows
\begin{equation}
{ U}\otimes { U'}={\bigoplus}_{k}(r_{k}, {V}_{k}),
\end{equation}
where $r_{k}$ and ${ V}_{k}$ are the representations of $\Gamma$
and $H$, respectively. The light chiral fermions in the
representation $r_{k}$ are determined by the zero modes of the
Dirac operator on $\Sigma$. The net number of chiral superfields
 is given by
\begin{eqnarray}
N_{r_{k}}-N_{r^{*}_{k}}=\chi(\Sigma,K^{1/2}_{\Sigma}\otimes
{\mathbf{V}_{k}}),
\end{eqnarray}
where $K_{\Sigma}$ is the  restriction of
canonical bundle on the curve $\Sigma$, and
$\mathbf{V}_{k}$ is the vector bundle whose sections 
transform in the representation ${ V}_{k}$ of 
the structure group $H$. 

In the F-theory model building, we are interested in the 
models where $G_{S'}$ is $U(1)'$, and
$H_S$ and $H_{S'}$ are respectively $U(1)$
and $U(1)'$. Then the vector bundles on $S$ and $S'$ 
are line bundles $L$ and $L'$. The adjoint representation
${\rm ad}G_{\Sigma}$ of $G_{\Sigma}$ is 
decomposed into a direct sum
of the irreducible representations under the group
$\Gamma_S \times U(1) \times U(1)'$ that can be
denoted as $\mathbf{(r_j, q_j, q'_j)}$
\begin{equation}
{\rm ad}G_{\Sigma}={\rm ad}(\Gamma_S)
\oplus {\rm ad}G_{S'}\oplus_{j}
\mathbf{(r_j, q_j, q_j')}~.~\,
\end{equation}
The numbers of chiral supefields  in the representation 
$\mathbf{(r_j, q_j, q'_j)}$ and their Hermitian conjugates
on the curve $\Sigma$ are given by 
\begin{eqnarray}
N_{\mathbf{(r_j, q_j, q'_j)}} ~=~ h^0 (\Sigma, \mathbf{V}_j) ~,~~~
N_{\mathbf{({\bar r}_j, -q_j, -q'_j)}} ~=~ h^1(\Sigma, \mathbf{V}_j)~,~\,
\end{eqnarray}
where 
\begin{eqnarray}
\mathbf{V}_j~=~ K^{1/2}_{\Sigma} \otimes
{L}_{\Sigma}^{q_{j}}\otimes {L'}_{\Sigma}^{q'_{j}} ~,~\,
\end{eqnarray}
where $K^{1/2}_{\Sigma}$, ${L}_{\Sigma}^{r_{j}}$ and
${L'}_{\Sigma}^{q'_{j}}$ are the restrictions of
canonical bundle $K_S$, line bundles $L$ and $L'$ on the curve
$\Sigma$, respectively. In particular, if the
volume of $S'$ is infinite, $G_{S'}=U(1)'$ is decoupled.
And then the index $\mathbf{q'_j}$ can be ignored.

Using Riemann-Roch theorem, we obtain the net number of 
chiral supefields in the representation $\mathbf{(r_j, q_j, q'_j)}$
\begin{eqnarray}
N_{\mathbf{(r_j, q_j, q'_j)}}-
N_{\mathbf{({\bar r}_j, -q_j, -q'_j)}}~=~ 1-g+c_1(\mathbf{V}_j) ~,~\,
\end{eqnarray}
where $g$ is the genus of the curve $\Sigma$, and $c_1$ means the first
Chern class.

Moreover, we can obtain the Yukawa couplings 
at the triple intersection of
three curves $\Sigma_i$, $\Sigma_j$ and $\Sigma_k$ where
 the gauge group or the singularity type is enhanced further.
To have the triple intersections, the corresponding
homology classes  $[\Sigma_i]$, $[\Sigma_j]$ and $[\Sigma_k]$
of the curves $\Sigma_i$, $\Sigma_j$ and $\Sigma_k$ must satisfy
the following conditions
\begin{eqnarray}
[\Sigma_i] \cdot [\Sigma_j] > 0 ~,~~~
[\Sigma_i] \cdot [\Sigma_k] > 0 ~,~~~
[\Sigma_j] \cdot [\Sigma_k] > 0 ~.~\,
\label{FTYK-Con}
\end{eqnarray}



The $SU(5)$ models, 
flipped $SU(5)\times U(1)_X$ models,
and $SU(3)_C\times SU(2)_L\times SU(2)_R\times U(1)_{B-L}$ models with 
additional vector-like particles have been constructed
previously~\cite{Beasley:2008dc, Beasley:2008kw, 
Font:2008id, Jiang:2009zza, Jiang:2009za, Li:2009cy, Li:2010dp}.
 However, the $SU(5)$ models with generic vector-like particles have 
not been studied systematically yet.
Thus, we shall construct the $SU(5)$ models
with additional vector-like particles in general here.
In such $SU(5)$ models, we introduce the  vector-like particles $YF'$
and ${\overline{YF}}'$, and $Yf'$
and ${\overline{Yf}'}$, whose quantum numbers under $SU(5)$ are
\begin{eqnarray}
YF'={\mathbf{10}}~,~
\overline{YF}'={\mathbf{\overline{10}}} ~;~  Yf'={\mathbf{5}}~,~
\overline{Yf}'={\mathbf{\overline{5}}} ~.~\,
\end{eqnarray}
Moreover,  the particle contents from the decompositions of
$YF'$, ${\overline{YF}}'$, $Yf'$, and ${\overline{Yf}}'$
under the SM gauge symmetry are
\begin{eqnarray}
&& YF' = (XQ, XU^c, XE^c)~,~ {\overline{YF}}'=(XQ^c, XU, XE)~,~\\
&& Yf'=(XD, XL^c)~,~ {\overline{Yf}}'= (XD^c, XL)~.~\,
\end{eqnarray}



Assuming that $S$ is a   $dP_8$ surface, 
we consider the observable gauge group $SU(5)$. 
On codimension-one curves that are  the intersections 
of the observable seven-branes and other
seven-branes, we obtain the
SM fermions, Higgs fields, and extra vector-like particles. To break the 
$SU(5)$ gauge symmetry down to
the $SU(3)_C\times SU(2)_L \times  U(1)_Y$ gauge symmetries, 
we turn on the $U(1)_Y$ flux on $S$ specified by the line bundle $L$. 
To obtain the SM fermions, Higgs fields and additional vector-like particles, 
we also turn on the $U(1)$ fluxes on the other seven-branes that intersect
with the observable seven-branes, and we specify these fluxes
 by the line bundle $L^{\prime n}$.

We take the line bundle $L=\mathcal{O}_{S}(E_{1}-E_{2})^{6/5}$.
Note that $\chi(S, L^{5/6})=0$, we do not have the vector-like particles
 on the bulk $S$. Moreover,
the curves with homology classes for the matter fields, Higgs fields 
and vector-like particles, and the gauge bundle assignments for 
each curve in the $SU(5)$ models are given 
in Table~\ref{F-SU5}. From this table, we obtain:
all the SM fermions are localized on the matter curves
$\Sigma_{F'}$ and $\Sigma_{\overline{f}'}$; 
the Higgs fields $H_u$ and $H_d$
 are localized on the
curves $\Sigma_{Hu}$, and $\Sigma_{Hd}$, respectively; 
and the vector-like particles
$YF'$, $\overline{YF}'$, $Yf'$, $\overline{Yf}'$,
$(XQ,~XQ^c)$, $(XU,~XU^c)$, $(XD,~XD^c)$, 
$(XL,~XL^c)$, and $(XE,~XE^c)$
are localized on the curves 
$\Sigma_{F'}$, $\Sigma_{\overline{F'}}$, $\Sigma_{f'} $, 
 $\Sigma_{\overline{f'}} $, $\Sigma_{XQ}$, $\Sigma_{XU}$,
$\Sigma_{XD}$, $\Sigma_{XL}$, and  $\Sigma_{XE}$,  respectively.
In addition, there exist singlets
from the intersections of the other seven-branes.
It is easy to check that we can realize the SM fermion
Yukawa coupling terms in our models. 
All the vector-like particles
can obtain masses by giving vacuum expectation values (VEVs)
to the SM singlets at the intersections of 
the other seven-branes.
Furthermore,  if we take the line bundle 
$L=\mathcal{O}_{S}(E_{1}-E_{2}+E_7-E_8)^{6/5}$.
we shall have one pair of vector-like particles
$(XY, ~XY^c)$ on the bulk $S$ because  $\chi(S, L^{5/6})=-1$.



\begin{table}[htb]
\begin{center}
\begin{tabular}{|c|c|c|c|c|c|}
\hline
 Fields & Curves  & ${\rm Class}$ & $g_{\Sigma}$ &
$L_{\Sigma}$ & $L_{\Sigma}^{\prime n}$\\\hline
$H_u$ & $\Sigma_{Hu}$ & $2H-E_{1}-E_3$ & $0$ &
$\mathcal{O}(1)^{6/5}$ & $\mathcal{O}(1)^{2/5}$\\\hline
$H_d$ & $\Sigma_{Hd}$ & $2H-E_{2}-E_3$ & $0$ &
$\mathcal{O}(-1)^{6/5}$ & $\mathcal{O}(-1)^{2/5}$\\\hline
$10_i+n\times XF'$ & $\Sigma_{F'}$ & $2H-E_{4}-E_6$ & $0$ 
& $\mathcal{O}(0)$ & 
$\mathcal{O}(3+n)$\\\hline
$n\times \overline{XF}'$ & $\Sigma_{\overline{F}'}$ & $2H-E_{5}-E_6$ & $0$ 
& $\mathcal{O}(0)$ & 
$\mathcal{O}(-n)$\\\hline
$\overline{5}_i+m\times \overline{Xf}'$ & 
$\Sigma_{\overline{f}'}$ & $H-E_7$ & $0$ 
& $\mathcal{O}(0)$ & 
$\mathcal{O}(-3-m)$\\\hline
$m\times Xf'$ & 
$\Sigma_{f'}$ & $H-E_8$ & $0$ 
& $\mathcal{O}(0)$ & 
$\mathcal{O}(m)$\\\hline
$(XQ, ~XQ^c)$ & $\Sigma_{XQ}$ & $3H-E_1-E_2 ~({\rm pinched})$ & $1$ &
 $\mathcal{O} (p_{12})^{6/5}$ & 
$\mathcal{O} (p_{12})^{-1/5}$ \\\hline
$(XU, ~XU^c)$ & $\Sigma_{XU}$ & $3H-E_1-E_2-E_3 ~({\rm pinched})$ & $1$ &
 $\mathcal{O} (p^3_{12})^{6/5}$ & 
$\mathcal{O} (p^3_{12})^{4/5}$ \\\hline
$(XD, ~XD^c)$ & $\Sigma_{XD}$ & $3H-E_1-E_2-E_4 ~({\rm pinched})$ & $1$ &
 $\mathcal{O} (p^4_{12})^{6/5}$ & 
$\mathcal{O} (p^4_{12})^{2/5}$ \\\hline
$(XL, ~XL^c)$ & $\Sigma_{XL}$ & $3H-E_1-E_5 ~({\rm pinched})$ & $1$ &
 $\mathcal{O} (p^5_{12})^{6/5}$ & 
$\mathcal{O} (p^5_{12})^{-3/5}$ \\\hline
$(XE, ~XE^c)$ & $\Sigma_{XE}$ & $3H-E_1-E_2-E_6 ~({\rm pinched})$ & $1$ &
 $\mathcal{O} (p^6_{12})^{6/5}$ & 
$\mathcal{O} (p^6_{12})^{-6/5}$ \\\hline
\end{tabular}
\end{center}
\caption{ The particle curves and 
gauge bundle assignments for each curve in the 
$SU(5)$ models from F-theory. Here 
 $i=1,~2,~3$. Moreover, $p_{12}=p_1-p_2$,
 $p_{12}^l = p_1^l-P_2^l$ for $l=3, ~4, ~5, ~6$,
and we denote the corresponding blowing up points as $p_1$, $p_2$,
 $p_1^l$, and $p_2^l$.}
\label{F-SU5}
\end{table}




\section{Gaugino Mass Relations and Their Indices}

First, let us briefly review the generalization of mSUGRA.
In four-dimensional GUTs with high-dimensional 
operators~\cite{Ellis:1985jn, Hill:1983xh, Shafi:1983gz, Ellis:1984bm, 
Drees:1985bx}, and F-theory $SU(5)$ 
models~\cite{Donagi:2008kj, Blumenhagen:2008aw} and 
$SU(3)_C \times SU(2)_L \times SU(2)_R \times U(1)_{B-L}$ 
models~\cite{Li:2009cy}, the
SM gauge kinetic functions are not unified at the GUT scale.
In general, the gaugino masses at the GUT scales can be
parametrized as follows~\cite{Li:2010xr}
\begin{eqnarray}
{{M_i}\over {\alpha_i}} &=& M^U_{1/2} + a_i M^{NU}_{1/2} ~,~\,
\label{GM-mSUGRA}
\end{eqnarray}
where $M^U_{1/2}$ and $M^{NU}_{1/2}$ are the universal and non-universal
gaugino masses at the GUT scale.
Thus, we define the index $k$ of the gaugino mass relation 
by the following equation~\cite{Li:2010xr}
\begin{eqnarray}
{{M_2}\over {\alpha_2}} - {{M_3}\over {\alpha_3}} 
~=~k \left( {{M_1}\over {\alpha_1}} 
- {{M_3}\over {\alpha_3}} \right) ~,~\,
\label{GMRelation}
\end{eqnarray}
where
\begin{eqnarray}
k & \equiv & {{a_2 -a_3}\over {a_1-a_3}} ~.~\,
\end{eqnarray}
Because $M_i/\alpha_i$ are renormalization scale invariant under 
one-loop RGE running and the two-loop RGE running 
effects are very small~\cite{Li:2010mr}, the gaugino mass relation 
in Eq.~(\ref{GMRelation}) can be preserved very well 
at low energy. Note that the gaugino masses
can be measured from the LHC and ILC 
experiments~\cite{Cho:2007qv, Barger:1999tn}, 
we can determine  $k$ at low energy.
 In addition, we have the following gauge coupling relation 
at the GUT scale
\begin{eqnarray}
{{1}\over {\alpha_2}} - {{1}\over {\alpha_3}} 
~=~k \left( {{1}\over {\alpha_1}} 
- {{1}\over {\alpha_3}} \right) ~.~\,
\label{GCRelation}
\end{eqnarray}
Thus, we can define the GUT scale via the above gauge coupling
relation. In short,
the index $k$  describes not only the gauge coupling
relation in Eq.~(\ref{GCRelation}) at the GUT scale, but also  
the gaugino mass relation in Eq.~(\ref{GMRelation}) 
which is exact from the
GUT scale to the electroweak scale at one loop.
Although $k$ is not well defined in the mSUGRA, in this paper,
we symbolically define the index $k$ for
the mSUGRA gaugino mass relation as $0/0$,
${\it i.e.}$, $k=0/0$ means the mSUGRA gaugino mass relation.

Interestingly, in the GMSB and AMSB, the gaugino masses are
 given by  Eq.~(\ref{GM-mSUGRA}) with $M^U_{1/2}=0$.
Thus, $M_i/(a_i\alpha_i)$ are
proportional to the same constant. And then we can define
their gaugino mass relations as follows 
\begin{eqnarray}
{{M_3}\over {a_3\alpha_3}} ~=~  {{M_2}\over {a_2\alpha_2}} 
~=~  {{M_1}\over {a_1\alpha_1}} ~.~\, 
\label{GAgmr}
\end{eqnarray}
Therefore, to present the gaugino mass relations in the
GMSB and AMSB, we only need to calculate $a_i$ in
the following.

\section{Gauge Mediated Supersymmetry Breaking}

First, let us consider the gaugino mass relations
and their indices in
the GMSB~\cite{gaugemediation}. 
In the messenger sector, we introduce a set
of the SM vector-like particles $\Phi_j$ and $\overline{\Phi}_j$.
To break supersymmetry, we introduce a chiral 
superfield $X$, whose F-term breaks supersymmetry. 
The messenger fields
couple to $X$ via the following superpotential
\begin{eqnarray}
W \supset \lambda_j X \overline{\Phi}_j \Phi_j ~,~\,  
\end{eqnarray}
where $\lambda_i$ are Yukawa couplings.
For simplicity, we assume that the scalar and auxiliary 
components of $X$ obtain VEVs
\begin{eqnarray}
\langle X \rangle &=& M + \theta^2 F ~.~\,  
\end{eqnarray}
Thus, the fermionic components of $\Phi_j$ and $\overline{\Phi}_j$
form Dirac fermions with masses $\lambda_j M$.
Denoting the superfields and their scalar components
 of $\Phi_j$ and $\overline{\Phi}_j$
in the same symbols, we obtain that  their scalar
components have the following squared-mass matrix in the
basis $(\Phi_j, ~\overline{\Phi}_j^{\dagger})$
\begin{eqnarray}
M^2~=~
\begin{pmatrix}
|\lambda_j M|^2 & - (\lambda_j F)^{\dagger} \cr - (\lambda_j F) & |\lambda_j M|^2
\end{pmatrix}~.~\,
\end{eqnarray}
Therefore, the scalar messenger
 mass eigenvetors are $(\Phi_j+\overline{\Phi}_j^{\dagger})/{\sqrt 2}$
and $(\Phi_j-\overline{\Phi}_j^{\dagger})/{\sqrt 2}$, and the
corresponding squared-mass eigenvalues are $(\lambda_j M)^2\pm \lambda_j F$.
The supersymmetry breaking, which is obvious in the messenger spectrum,
is communicated to the SM sector via the gauge interactions of 
$\Phi_j$ and $\overline{\Phi}_j$. And then we obtain the gaugino masses
at one loop as follows
\begin{eqnarray}
{{M_i}\over {\alpha_i}} &=& {1\over {4\pi}} {F\over M} \sum_j 
n_i(\Phi_j) g(x_j)~,~\,  
\end{eqnarray}
where $n_i(\Phi_j)$ is the sum of Dynkin indices for the vector-like particles
 $\Phi_j$ and $\overline{\Phi}_j$, $x_j=|F/(\lambda_j M^2)|$, and
\begin{eqnarray}
g(x)~ =~ {1\over {x^2}} \left[ (1+x){\rm ln}(1+x)+(1-x){\rm ln}(1-x) \right]  ~.~\,  
\end{eqnarray}
Approximately, we have the expansion of $g(x)$ as follows
\begin{eqnarray}
g(x)~ =~ 1+ {{x^2}\over 6} +   {{x^4}\over {15}} + {{x^6}\over {28} }
+\cdots ~.~\,  
\end{eqnarray}
Because the squared-masses for the messenger fields must be positive,
we obtain $0 \le x_j \le 1$. Also, $g(x)$ is a monotonically increasing 
function from $g(0)=1$ to $g(1)=1.386$.
Therefore, in the GMSB, we have
\begin{eqnarray}
a_i &=& \sum_j  n_i(\Phi_j) g(x_j) ~.~\,  
\end{eqnarray}
In particular, 
if all the messenger fields have the same Yukawa couplings to $X$,
${\it i.e}$, $\lambda_j$ are the same, we have 
\begin{eqnarray}
a_i &=& \sum_j  n_i(\Phi_j)  ~.~\,  
\label{Ai-GAMSB}
\end{eqnarray}
Moreover, if the messenger fields are heavier than $10^7$ GeV
and their Yukawa couplings to $X$ are about order one for naturalness, 
we obtain $x_j \le 0.1$, and then $g(x_j)\simeq 1$. So we have 
\begin{eqnarray}
a_i & \simeq & \sum_j  n_i(\Phi_j)  ~.~\,  
\end{eqnarray}

To preserve the gauge coupling unification in GUTs, we usually
assume that the vector-like messengers form complete $SU(5)$
multiplets, for example, $({\mathbf{5}}, ~{\mathbf{\overline{5}}})$.
In general, the messengers do not need to form complete
 $SU(5)$ multiplets. To achieve the gauge coupling unification, 
we can introduce extra vector-like particles
around the same messeger scale, which do not couple to supersymmetry
breaking chiral superfield $X$. For example,
assuming that we have the vector-like messenger fields $(XD, ~XD^c)$ 
(or $(XL, ~XL^c)$), we introduce the vector-like particles  
$(XL, ~XL^c)$ (or $(XD, ~XD^c)$) at the messenger scale
so that the gauge coupling unification can be preserved.
In GUTs from orbifold constructions, 
intersecting D-brane model building on Type II orientifolds,
M-theory on $S^1/Z_2$ with Calabi-Yau compactifications, 
and F-theory model building, $(XD, ~XD^c)$ and $(XL, ~XL^c)$
do not need to arise from the same GUT multiplets
since the zero modes of their triplet partners and doublet 
partners can be projected out, respectively. Thus, we can 
realize such scenarios with some fine-tuning. Interestingly,
in the flipped $SU(5)\times U(1)_X$ models, we do not need
to fine-tune the mass scales for the vector-like particles
due to the two-step gauge coupling unification.


\begin{table}[htb]
\begin{center}
\begin{tabular}{|c|c|c|c|||c|c|c|c|}
\hline
~${\rm Cases }$~ &
${\rm Messengers }$ & $(n_1, n_2, n_3)$ & $k$ & 
~${\rm Cases }$~ &
${\rm Messengers }$ & $(n_1, n_2, n_3)$ & $k$ \\\hline
$(1)$ &
~$(XQ, ~XQ^c)$~ & $(1/5, 3, 2)$ & ~$-5/9$~ & 
$(2)$ &
~$(XU, ~XU^c)$~ & $(8/5, 0, 1)$ & ~$-5/3$ ~  \\ \hline
$(3)$ &
$(XD, ~XD^c)$ & $(2/5, 0, 1)$ & $5/3$ & 
$(4)$ &
$(XL, ~XL^c)$ & $(3/5, 1, 0)$ & $5/3$  \\ \hline
$(5)$ &
$(XE, ~XE^c)$ & $(6/5, 0, 0)$ & $0$ & 
$(6)$ &
$(XY, ~XY^c)$ & $(5, 3, 2)$ & $1/3$  \\ \hline
$(7)$ &
$XG$ &  $(0, 0, 3)$ & $1$ & 
$(8)$ &
$XW$ &  $(0, 2, 0)$ & $\infty$ \\ \hline
$(9)$ &
$(XT, ~XT^c)$ & ~$(18/5, 4, 0)$~ & $10/9$ & 
$(10)$ &
$(XS, ~XS^c)$ & ~$(16/5, 0, 5)$ ~& $25/9$  \\ \hline
$(11)$ &
$(XQ, ~XQ^c)$  & $(7/5, 3, 2)$ & $-5/3$ & 
$(12)$ &
$(XU, ~XU^c)$ & $(14/5, 0, 1)$ & $-5/9$ 
 \\ 
 & $(XE, ~XE^c)$  &    & & & $(XE, ~XE^c)$ &  
&   \\ \hline
$(13)$ &
$XG$ &  $(0, 2, 3)$ & $1/3$ & 
$(14)$ & $(XT, ~XT^c)$ & $(34/5, 4, 5)$ & $- 5/9$  \\ 
 & $XW$ &     &  & & 
$(XS, ~XS^c)$ &  &    \\ \hline
$(15)$ &
$(\mathbf{5}, \mathbf{\overline{5}})$ &  $(1, 1, 1)$ & $0/0$ & 
$(16)$ &
$(\mathbf{10}, \mathbf{\overline{10}})$  &  $(3, 3, 3)$ & $0/0$  \\ \hline
$(17)$ &
$(\mathbf{15}, \mathbf{\overline{15}})$ &  $(7, 7, 7)$ & $0/0$ & 
$(18)$ &
$\mathbf{24}$  &  $(5, 5, 5)$ & $0/0$  \\ \hline
\end{tabular}
\end{center}
\caption{  The $n_i(\Phi)$ for the messenger fields
and the corresponding indices $k$ of the gaugino mass relations
in $SU(5)$ models. }
\label{GUT-SUV}
\end{table}


 To calculate the parameters $a_i$ and indices $k$ for the gaugino mass 
relations,  we assume for simplicity that either all the messenger 
fields have the same Yukawa couplings to $X$, or the  messenger fields 
are  heavier than $10^7$ GeV, and then, the parameters $a_i$ are given
by Eq.~(\ref{Ai-GAMSB}). Thus, we only need to present
the Dynkin indices $n_i$ for the messenger fields. 
We emphasize that with the gaugino mass
relations and their indices $k$, we may probe the messenger
fields at the intermediate scale. With various messenger fields,
we shall consider $SU(5)$ models, 
flipped $SU(5)\times U(1)_X$ models with $SO(10)$
origin, and Pati-Salam Models with $SO(10)$
origin in the following: \\

(i) $SU(5)$ Models \\

In Table~\ref{GUT-SUV}, we present
the $n_i(\Phi)$ for the messenger fields
and the corresponding indices $k$ of the gaugino mass relations
in $SU(5)$ models. We can construct 
orbifold $SU(5)$ models with vector-like particles
in the Cases $(1)$, $(3)$, $(4)$, $(6)$, $(12)$, $(13)$, $(14)$, 
$(15)$, $(16)$, $(17)$, and  $(18)$ in Table~\ref{GUT-SUV}.
Here, the Cases $(15)$, $(16)$, $(17)$, and  $(18)$ can
be considered as the combinations of two Cases, Cases $(3)$ and $(4)$, 
Cases $(1)$ and $(12)$, Cases $(1)$ and $(14)$,
and Cases $(6)$ and $(13)$, respectively.
Assuming the superpotential between the messenger
fields and $X$ is on the D3-brane at $y=\pi R/2$ where
only the SM gauge symmetries is preserved, we can construct 
orbifold $SU(5)$ models with vector-like particles
in the rest Cases in Table~\ref{GUT-SUV}, {\it i.e.}, the Cases
$(2)$, $(5)$, $(7)$, $(8)$, $(9)$, $(10)$, and $(11)$.
Moreover,  in the F-theory $SU(5)$ models, we can construct the
 $SU(5)$ models with vector-like particles
in the Cases $(1)$, $(2)$, $(3)$,  $(4)$, $(5)$, $(6)$, $(11)$, $(12)$,   
$(15)$, and $(16)$ in Table~\ref{GUT-SUV}.
In addition, for the Cases  $(2)$, $(3)$, $(4)$,  
$(9)$, $(10)$,   $(12)$, and $(13)$, there are one
massless gaugino, and in the Cases $(5)$,  $(7)$, and $(8)$,  
there are two massless
gauginos. Thus, each of these Cases can not be consistent
with the low-energy phenomenological constraints. To give masses
to all the SM gauginos, we can combine the different Cases,
and the corresponding indices can be calculated similarly.
For example, we can add the messenger fields 
$(\mathbf{5}, \mathbf{\overline{5}})$ for each of these
Cases. Then the Dynkin indices for the messenger fields
increase by one, ${\it i.e.}$, we change $n_i$ to $n_i+1$
for each of these Cases in Table~\ref{GUT-SUV}.
Interestingly, the indices $k$ are the same as
those in  Table~\ref{GUT-SUV} since
$(\mathbf{5}, \mathbf{\overline{5}})$ form complete 
$SU(5)$ representations.
Also, some interesting combinations of the different 
Cases will be studied 
in the flipped  $SU(5)\times U(1)_X$ models and
the Pati-Salam $SU(4)_C\times SU(2)_L \times SU(2)_R$ models in the
following.
Furthermore, we emphasize that we do have the mSUGRA gaugino
mass relation if the 
messenger fields form the complete $SU(5)$ representations.
Also, if two sets of the messenger fields form complete $SU(5)$
representations, we can show that the indices $k$ for these two sets 
of the messenger fields are the same. For example, the messenger fields
$(XD,~XD^c)$ and $(XL, XL^c)$ have the same index $k=5/3$. \\


(ii) Flipped $SU(5)\times U(1)_X$  Models \\

In Table~\ref{GUT-FSUV}, we present
the $n_i(\Phi)$ for the messenger fields
and the corresponding indices $k$ of the gaugino mass relations
in flipped $SU(5)\times U(1)_X$ models. We can construct the
orbifold $SO(10)$ models  with vector-like particles
in the Cases $(1)$,  $(4)$, $(5)$, $(6)$, $(8)$, and $(11)$ 
in Table~\ref{GUT-FSUV}
where the $SO(10)$ gauge symmetry is broken down to the
flipped $SU(5)\times U(1)_X$ gauge symmetries.
Assuming the superpotential between the messenger
fields and $X$ is on the D3-brane at $y=\pi R/2$ where
only the $SU(5)\times U(1)_X$ gauge symmetries is preserved, 
we can construct the
orbifold $SO(10)$ models with vector-like particles
in the rest Cases in Table~\ref{GUT-FSUV}, 
{\it i.e.}, the Cases
$(2)$, $(3)$,  $(7)$, $(9)$, $(10)$, and $(12)$.
Moreover, in the F-theory $SO(10)$ models where
the gauge symmetry is broken down to the
flipped $SU(5)\times U(1)_X$  gauge symmetries
by turning on the $ U(1)_X$ flux, we can construct the
flipped  $SU(5)\times U(1)_X$ models with vector-like particles
in all the Cases in the Table~\ref{GUT-FSUV} 
except the Case $(5)$~\cite{Jiang:2009zza, Jiang:2009za}.
Interestingly, the indices $k$ for the gaugino mass relations are zero
for all the Cases except the Case $(4)$ with messenger
fields $(Xh, ~\overline{Xh})$. For the Case $(4)$,
we obtain the mSUGRA gaugino mass
relation. In addition,  we have two massless
gauginos in the Case $(3)$, so it  can not be consistent with the
low-energy phenomenological constraints by itself.
Furthermore,  for the Cases $(1)$, $(4)$,  $(5)$, $(7)$, 
  $(10)$, and $(11)$, we can realize the gauge coupling unification
naturally. While for
the Cases $(2)$, $(3)$,  $(6)$, $(8)$, $(9)$, and $(12)$,
we can achieve the gauge coupling unification  in the testable 
flipped $SU(5)\times U(1)_X$ models due to the two-step
gauge coupling unification. \\




\begin{table}[htb]
\begin{center}
\begin{tabular}{|c|c|c|c||c|c|c|c|}
\hline
~${\rm Cases }$~ &
${\rm Messengers }$ & $(n_1, n_2, n_3)$ & ~$k$~ & 
~${\rm Cases }$~ &
${\rm Messengers }$ & $(n_1, n_2, n_3)$ & ~$k$~ 
\\\hline
$(1)$ &
$(XF, ~\overline{XF})$ & ~$(3/5, 3, 3)$~ & $0$ & 
$(2)$ &
$(Xf, ~\overline{Xf})$ & $(11/5, 1, 1)$ & $0$  \\ \hline
$(3)$ &
$(Xl, ~\overline{Xl})$ & $(6/5, 0, 0)$ & $0$ &  
$(4)$ &
$(Xh, ~\overline{Xh})$ & $(1, 1, 1)$ &~ $0/0$~  \\ \hline
$(5)$ &
~$(XGW, XN)$~ & $(1/5, 5, 5)$ & $0$ & 
$(6)$ &
~$(XX, ~\overline{XX})$~ &~ $(39/5, 3, 3)$~ & $0$  \\ \hline
$(7)$ &
$(XF, ~\overline{XF})$  & $(9/5, 3, 3)$ & $0$ & 
$(8)$ &
$(Xf, ~\overline{Xf})$  & $(17/5, 1, 1)$ & $0$  \\ 
 & $(Xl, ~\overline{Xl})$  &   &  &
 & $(Xl, ~\overline{Xl})$  &   & \\ \hline
$(9)$ &
$(Xh, ~\overline{Xh})$ & $(11/5, 1, 1)$ & $0$ & 
$(10)$ &
$(XF, ~\overline{XF})$ & $(14/5, 4, 4)$ & $0$  \\  
 & $(Xl, ~\overline{Xl})$ &  &  &  
 & $(Xf, ~\overline{Xf})$ &  
& \\  \hline
$(11)$ &
$(XF, ~\overline{XF})$ & $(8/5, 4, 4)$ & $0$ & 
$(12)$ &
$(Xf, ~\overline{Xf})$ & $(16/5, 2, 2)$ & $0$  \\ 
 & $(Xh, ~\overline{Xh})$ & &   &
 & $(Xh, ~\overline{Xh})$ &    & \\  \hline
\end{tabular}
\end{center}
\caption{  The $n_i(\Phi)$ for the messenger fields
and the corresponding indices $k$ of the gaugino mass relations
in flipped $SU(5)\times U(1)_X$ models.}
\label{GUT-FSUV}
\end{table}





(iii)   Pati-Salam $SU(4)_C\times SU(2)_L \times SU(2)_R$ Models \\

In Table~\ref{GUT-PS}, we present
the $n_i(\Phi)$ for the messenger fields
and the corresponding indices $k$ of the gaugino mass relations
in Pati-Salam $SU(4)_C\times SU(2)_L \times SU(2)_R$ models. 
We can construct the
orbifold $SO(10)$ models  with vector-like particles
in all the Cases in Table~\ref{GUT-PS}
where the $SO(10)$ gauge symmetry is broken down to the
 Pati-Salam $SU(4)_C\times SU(2)_L \times SU(2)_R$ gauge symmetries.
Moreover, in F-theory $SO(10)$ models 
where the $SO(10)$ gauge symmetry is broken down to
the $SU(3)_C \times SU(2)_L \times SU(2)_R \times U(1)_{B-L}$
gauge symmetries by turning on the $U(1)_{B-L}$ 
flux~\cite{Font:2008id, Li:2009cy},
 we can construct the $SU(3)_C \times SU(2)_L \times SU(2)_R \times U(1)_{B-L}$
 models with vector-like particles
in all the Cases in the Table~\ref{GUT-PS} except the Case $(5)$~\cite{Li:2009cy}.
In addition, in the Cases $(2)$, $(3)$,  $(4)$, and $(8)$,  there
are one massless gaugino, and then each of them is not consistent
with the low-energy phenomenological constraints by itself. 
We can solve the problem by combining the different Cases, and
some combinations of the different simple Cases are 
 given in Table~\ref{GUT-PS} as well.





\begin{table}[htb]
\begin{center}
\begin{tabular}{|c|c|c|c||c|c|c|c|}
\hline
~${\rm Cases }$~ &
${\rm Messengers }$ & $(n_1, n_2, n_3)$ & $k$ & 
~${\rm Cases }$~ &
${\rm Messengers }$ & $(n_1, n_2, n_3)$ & $k$ 
\\\hline
$(1)$ &
~$(XFL, ~\overline{XFL})$~ & $(4/5, 4, 2)$ &~ $-5/3$~  & 
$(2)$ &
~$(XFR, ~\overline{XFR})$~ & $(16/5, 0, 2)$ & ~$-5/3$~  \\ \hline
$(3)$ &
$XD\overline{D}$ & $(2/5, 0, 1)$ & $5/3$ & 
$(4)$ &
$XL\overline{L}$ & $(3/5, 1, 0)$ & $5/3$   \\ \hline
$(5)$ &
$(XG4, XWL)$ & ~$(14/5, 2, 4)$~ & $5/3$ & 
$(6)$ &
$XZ$ & ~$(26/5, 6, 4)$~ & $5/3$  \\ 
 & $ XWR$ & &    &
 & &  &   \\ \hline
$(7)$ &
$(XFL, ~\overline{XFL})$ & $(6/5, 4, 3)$ & ~$-5/9$~ &  
$(8)$ &
$(XFR, ~\overline{XFR})$  & $(18/5, 0, 3)$ & $-5$   \\ 
 & $XD\overline{D}$  &    &  &
 & $XD\overline{D}$  &   & \\ \hline
$(9)$ &
$(XFL, ~\overline{XFL})$ & $(7/5, 5, 2)$ & $-5$ & 
$(10)$ &
$(XFR, ~\overline{XFR})$  & $(19/5, 1, 2)$ & ~$-5/9$ ~ \\ 
 & $XL\overline{L}$   &    &  &
 & $XL\overline{L}$   &   & \\ \hline
\end{tabular}
\end{center}
\caption{  The $n_i(\Phi)$ for the messenger fields
and the corresponding indices $k$ of the gaugino mass relations
in Pati-Salam $SU(4)_C \times SU(2)_L \times SU(2)_R$ models.}
\label{GUT-PS}
\end{table}



\section{Anomaly Mediated Supersymmetry Breaking}

We first briefly review the AMSB~\cite{anomalymediation,
UVI-AMSB, D-AMSB}. The supergravity Lagrangian
can be obtained from a local superconformal field theory
by a gauge fixing of extra symmetries, which can be done
by setting the values of the components of a chiral
compensator field $C$. Thus, $C$ couples to the conformal
symmetry violation, {\it i.e.}, all the dimensionful parameters
including the renormalization scale $\mu$. To have the
canonical normalization for the gravity kinetic terms,
we determine the scalar component of $C$. To cancel the 
cosmological constant after supersymmetry breaking in the
hidden sector, we give a non-zero VEV to the auxiliary
component $F^C$ of $C$, which is the only source of supersymmetry
breaking. With $\langle C \rangle = M_C + \theta^2 F^C$,
 we obtain the gravitino mass $m_{3/2}=F^C/M_C$.
To suppress the supergravity contributions to
the supersymmetry breaking soft terms, we assume
the sequestering between the observable and hidden sectors 
for simplicity. This can be realized naturally
in the five-dimensional brane world scenario where
the observable and hidden sectors are confined on the
different branes~\cite{sequestering}, 
or in the models where the contact
terms between the observable and hidden sectors are 
suppressed dynamically by a conformal 
sector~\cite{conformalsequestering}.

In this paper, we concentrate on the gaugino masses.
The relevant Lagrangian is
\begin{eqnarray}
{\cal L} \supset \int d^2\theta {1\over {2g^2}} 
{\rm Tr}\left[W^{\alpha} W_{\alpha}\right] + {\rm H.C.} ~,~\,  
\end{eqnarray}
where $W^{\alpha}$ is the field strength of the
vector superfield. Because the compensator $C$ couples to 
the renormalization scale $\mu$, there are additional
contributions at quantum level. Then we have 
\begin{eqnarray}
{\cal L} \supset \int d^2\theta {1\over\displaystyle {2g^2
\left({\mu \over C}\right)}} 
{\rm Tr}\left[W^{\alpha} W_{\alpha}\right] + {\rm H.C.} ~.~\,  
\end{eqnarray}
Thus, we obtain the SM gaugino masses
\begin{eqnarray}
{{M_i}\over {\alpha_i}}~=~{{b_i}\over {4\pi}} {{F^C}\over {M_C}} ~,~\, 
\label{AMSB-GM} 
\end{eqnarray}
where $b_3$, $b_2$, and $b_1$ are the one-loop beta functions
for $SU(3)_C$, $SU(2)_L$, and $U(1)_Y$, respectively.
In particular,
if there are vector-like particles at the intermediate scales
which do not mediate supersymmetry breaking,
we emphasize that these vector-like particles will not
affect the low-energy gaugino masses in the AMSB 
after they are integrated out~\cite{Choi:2007ka}.

Moreover,
although AMSB can solve the flavour changing neutral current
problem, the minimal AMSB is excluded since the squared slepton
masses are negative and then the electromagnetism
will be broken. In this paper, we consider two solutions:
(1) UV insensitive anomaly mediation~\cite{UVI-AMSB}; 
(2) Deflected anomaly
mediation~\cite{D-AMSB}.

\subsection{UV Insensitive Anomaly Mediation}

In the UV insensitive anomaly mediation~\cite{UVI-AMSB}, the $U(1)$ D-terms
 contribute to the slepton masses, and then
  the squared slepton masses  can be positive.
In particular, the $U(1)$ symmetries can be $U(1)_Y$ and
$U(1)_{B-L}$ so that we only need to introduce three right-handed
neutrinos  to 
cancel the  $U(1)_{B-L}$ gauge anomalies. Interestingly, the gaugino
masses are still given by Eq.~(\ref{AMSB-GM}). Thus, we obtain
\begin{eqnarray}
a_i~=~b_i ~.~\, 
\end{eqnarray}

We shall consider the $SU(5)$ and flipped $SU(5)\times U(1)_X$
models with TeV-scale vector-like particles. 
To achieve the one-step gauge coupling
unification, we emphasize that the discussions for the
Pati-Salam $SU(4)_C\times SU(2)_L\times SU(2)_R$ models are 
similar to those in the $SU(5)$ models. Thus, we will not 
consider the Pati-Salam models here for simplicity.
In $SU(5)$ models, to achieve the gauge coupling unification,
we consider the TeV-scale vector-like particles that form
complete $SU(5)$ representations. In Table~\ref{SUV-AM},
we present the parameters 
$a_i$ and the indices $k$ of the gaugino mass
relations  in the $SU(5)$ models without
and with TeV-scale vector-like particles.
Especially, the indices $k$ are equal to $5/12$
for all these Cases.
In addition, we present  the parameters  $a_i$ and the indices $k$ of 
the gaugino mass relations in Table~\ref{FSUV-AM} 
in the  flipped $SU(5)\times U(1)_X$ models with 
TeV-scale vector-like particles. These vector-like
particles also form complete $SU(5)\times U(1)_X$ representations. 
For the Cases $(1)$, $(4)$, $(5)$, $(8)$, and $(9)$, 
we can have the gauge coupling unification naturally. 
However, for the Cases $(2)$, $(3)$, $(6)$, and $(7)$,
 we should introduce
the vector-like particles $(XF,~\overline{XF})$ at the intermediate
scale $10^8$ GeV or smaller
so that we can obtain the  gauge coupling unification.

Furthermore, for the Cases $(4)$ and  $(6)$ in the $SU(5)$ models
and the Cases $(1)$ and  $(5)$ in the flipped $SU(5)\times U(1)_X$
models,  gluino is massless. This problem can be solved
elegantly in the deflected AMSB in the next subsection. Also,
for the Cases  $(5)$ and  $(7)$ in the $SU(5)$ models
and the Cases $(8)$ and  $(9)$ in the flipped $SU(5)\times U(1)_X$
models, we emphasize that the masses of the 
vector-like particles may need to be about 20 TeV or 
larger so that
we can avoid the Landau pole problem for the strong 
coupling~\cite{Jiang:2009za, Jiang:2006hf}.
Thus, we can not test these models
 at the LHC since we may have 10 TeV scale
supersymmetry breaking.



\begin{table}[htb]
\begin{center}
\begin{tabular}{|c|c|c|c||c|c|c|c|}
\hline
~${\rm Case}$~  & ~${\rm New~Particles}$~ & ($a_1$,  $a_2$, $a_3$)   &  $k$ &
~${\rm Case}$~  & ~${\rm New~Particles}$~ & ($a_1$,  $a_2$, $a_3$)   &  $k$
 \\
\hline
(1) &
${\rm No}$ & ~ $(33/5, 1, -3)$ ~& ~$5/12$~ & 
(2) & $(\mathbf{5}, \mathbf{\overline{5}})$ & ~$(38/5, 2, -2)$~ & ~$5/12$~ \\ \hline
(3) & $2 \times (\mathbf{5}, \mathbf{\overline{5}})$
 &  $(43/5, 3, -1)$ & $5/12$ & 
(4) & $3\times (\mathbf{5}, \mathbf{\overline{5}})$ & $(48/5, 4, 0)$ & $5/12$ \\ \hline
(5) & $4 \times (\mathbf{5}, \mathbf{\overline{5}})$
 &  $(53/5, 5, 1)$ & $5/12$ & 
(6) & $ (\mathbf{10}, \mathbf{\overline{10}})$ & $(48/5, 4, 0)$ & $5/12$ \\ \hline
(7) & ~$ (\mathbf{5}, \mathbf{\overline{5}})$, 
$ (\mathbf{10}, \mathbf{\overline{10}})$~
 &  $(53/5, 5, 1)$ & $5/12$  & & & & \\  \hline
\end{tabular}
\end{center}
\caption{The parameters $a_i$ and the indices $k$ for the UV insensitive 
AMSB in the $SU(5)$ models without and with additional
vector-like particles.}
\label{SUV-AM}
\end{table}





\begin{table}[htb]
\begin{center}
\begin{tabular}{|c|c|c|c||c|c|c|c|}
\hline
~${\rm Case}$~  & ${\rm New~Particles}$ & ($a_1$,  $a_2$, $a_3$)   &  $k$ &
~${\rm Case}$ ~ & ${\rm New~Particles}$ & ($a_1$,  $a_2$, $a_3$)   &  $k$
 \\ \hline
(1) &
$(XF, \overline{XF})$ &  $(36/5, 4, 0)$ & $5/9$ & 
(2) & $(Xf, \overline{Xf})$ & $(44/5, 2, -2)$ & ~$10/27$~ \\ \hline
(3) &
$(Xl, \overline{Xl})$ 
 & ~ $(39/5, 1, -3)$ ~& ~$10/27$~ & 
(4) & $(Xh, \overline{Xh})$ & ~$(38/5, 2, -2)$ ~& $5/12$ \\ \hline
(5) & ~$(XF, \overline{XF})$~
& $(42/5, 4, 0)$ & $10/21$ &
(6) & $(Xf, \overline{Xf})$
& $(49/5, 3, -1)$ & $10/27$ \\
 & $(Xl, \overline{Xl})$ 
&  &  &
 &$(Xh, \overline{Xh})$ 
&  &  \\ \hline
(7) & $(Xl, \overline{Xl})$
&  $(44/5, 2, -2)$ & $10/27$ &
(8) &
~$(XF, \overline{XF})$~
 &  $(47/5, 5, 1)$ & $10/21$ \\ 
 & $(Xh, \overline{Xh})$ &   &  & &
~$(Xf, \overline{Xf})$ ~ &   &  \\ \hline
(9) & ~$(XF, \overline{XF})$~
& $(41/5, 5, 1)$ & $5/9$ & & & & \\
 & ~$(Xh, \overline{Xh})$ ~
&  &  & & & &
\\ \hline
\end{tabular}
\end{center}
\caption{The parameters $a_i$ and the indices $k$ 
for the UV insensitive AMSB in the 
flipped $SU(5)\times U(1)_X$ models with additional
vector-like particles.}
\label{FSUV-AM}
\end{table}



\subsection{Deflected Anomaly Mediation}

In the deflected anomaly mediation~\cite{D-AMSB}, similar to the
GMSB, we introduce  a chiral superfield $X$  and a set
of the SM vector-like particles $\Phi_j$ and $\overline{\Phi}_j$.
The superpotential is
\begin{eqnarray}
W \supset \lambda_j X \overline{\Phi}_j \Phi_j 
+ M_* ^{3-p} X^p ~,~\,  
\end{eqnarray}
where $p\not=3$, and $M_*$ is a model-dependent mass parameter.
The chiral compensator $C$ couples to $X$ at tree level 
by the scale non-invariant term $ M_* ^{3-p} X^p$, and
then the VEVs of $X$ can be fixed. It was shown that
$X$ is stabilized at $\langle X \rangle >> m_{3/2}$ for
$M_* >> m_{3/2}$ if $p> 3$ or $p<0$ as follows
\begin{eqnarray}
\langle X \rangle~=~M_X +\theta^2 F^X ~,~\,  
\end{eqnarray}
where
\begin{eqnarray}
M_X \simeq m_{3/2}^{1/(p-2)} M_*^{(p-3)/(p-2)}~,~~
{{F^X}\over {M_X}} ~=~ -{2\over {p-1}} {{F^C}\over {M_C}} ~.~\,  
\end{eqnarray}
In addition, even without the term $ M_* ^{3-p} X^p$ in
the superpotential, $X$ can still be stabilized by the
radiative corrections to its K\"ahler potential, and then
we have
\begin{eqnarray}
{{F^X}\over {M_X}} ~\simeq~ - {{F^C}\over {M_C}} ~.~\,  
\end{eqnarray}
Thus, the contributions to the supersymmetry breaking soft 
masses from gauge mediation are comparable to those from
 anomaly mediation, and then we can solve
the tachyonic slepton problem in the AMSB.
Moreover, we obtain the gaugino masses at the TeV scale 
\begin{eqnarray}
{{M_i}\over {\alpha_i}}~=~{{1}\over {4\pi}} 
\left( b_i + {2\over {p-1}} \sum_j n_i(\Phi_j) g(x_j)  \right)
{{F^C}\over {M_C}} ~.~\, 
\label{AMSB-DAM} 
\end{eqnarray}
Thus, we have 
\begin{eqnarray}
a_i~=~b_i + {2\over {p-1}} \sum_j n_i(\Phi_j) g(x_j) ~.~\, 
\end{eqnarray}
If the messenger fields are heavier than $10^7$ GeV
and their Yukawa couplings to $X$ are about order one,
 we obtain
\begin{eqnarray}
a_i~\simeq~b_i + {2\over {p-1}} \sum_j n_i(\Phi_j)  ~.~\, 
\end{eqnarray}
Thus, choosing the possible value for $p$ and introducing
the TeV-scale vector-like particles, we can calculate
the parameters $a_i$ and the 
indices $k$ of the gaugino mass relations.

To probe the messenger fields in the deflected anomaly mediation,
we should define a new index $k'$ for the gaugino mass relations.
In the supersymmetric SM, we have 
\begin{eqnarray}
b_1~=~{{33}\over 5}~,~~ b_2~=~ 1~,~~b_3=-3  ~.~\, 
\end{eqnarray}
Thus, $b_1$ and $b_2$ will aways be positive even if we introduce
the vector-like particles at the TeV scale. Therefore, 
for $b_3\not=0$, we define the new index $k'$ as follows
\begin{eqnarray}
k'~\equiv~{\displaystyle {{{b_1 b_3 {{M_2}\over {\alpha_2}}}
- {b_1 b_2 {{M_3}\over {\alpha_3}}}}} \over\displaystyle
{{{b_2 b_3 {{M_1}\over {\alpha_1}}}
-  {b_1 b_2 {{M_3}\over {\alpha_3}}}}}} 
~=~ {\displaystyle {b_1 b_3 \sum_j n_2(\Phi_j) g(x_j)
-  b_1 b_2\sum_j n_3(\Phi_j) g(x_j)}
\over\displaystyle {b_2 b_3 \sum_j n_1(\Phi_j) g(x_j)
-  b_1 b_2 \sum_j n_3(\Phi_j) g(x_j) }}  ~.~\, 
\end{eqnarray}
And for $b_3=0$, we define the new index $k'$ as follows
\begin{eqnarray}
k'~\equiv~{\displaystyle {{{b_1  {{M_2}\over {\alpha_2}}}
- { b_2 {{M_1}\over {\alpha_1}}}}} \over\displaystyle
{{ { {{M_3}\over {\alpha_3}}}}}} 
~=~ {\displaystyle {b_1  \sum_j n_2(\Phi_j) g(x_j)
-   b_2\sum_j n_1(\Phi_j) g(x_j)}
\over\displaystyle { \sum_j n_3(\Phi_j) g(x_j) }}  ~.~\, 
\end{eqnarray}

Assuming that the messenger fields are heavier than $10^7$ GeV
and their Yukawa couplings to $X$ are about order one,
we consider the $SU(5)$ models,
the flipped $SU(5)$ models,  the Pati-Salam Models,
and the other possible models in the following: \\




\begin{table}[htb]
\begin{center}
\begin{tabular}{|c|c|c|c||c|c||c|c|}
\hline
~${\rm Cases }$~ &
~${\rm Messengers }$~ & ~$(a_1^0, ~a_2^0, ~a_3^0)$~  & ~$k_0$~ & 
~$(a_1^1, ~a_2^1, ~a_3^1)$~  & ~$k_1$~ &  
~$(a_1^2, ~a_2^2, ~a_3^2)$~  & ~$k_2$~ \\\hline
$(1)$ &
$(XQ, ~XQ^c)$ & $({\displaystyle 101 \over\displaystyle 15}, 
3, -{\displaystyle 5 \over\displaystyle 3})$ 
& ${\displaystyle 5 \over\displaystyle 9}$ 
& $({\displaystyle 116 \over\displaystyle 15}, 
4, -{\displaystyle 2 \over\displaystyle 3})$ 
 & ${\displaystyle 5 \over\displaystyle 9}$ 
 & $({\displaystyle 146 \over\displaystyle 15}, 
6, {\displaystyle 4 \over\displaystyle 3}) $ &
${\displaystyle 5 \over\displaystyle 9}$  \\ \hline
$(2)$ &
$(XU, ~XU^c)$ & $({\displaystyle 23 \over\displaystyle 3}, 
1, -{\displaystyle 7 \over\displaystyle 3})$ 
& ${\displaystyle 1 \over\displaystyle 3}$ & 
$({\displaystyle 26 \over\displaystyle 3}, 2, 
-{\displaystyle 4 \over\displaystyle 3})$
& ${\displaystyle 1 \over\displaystyle 3}$

 & $({\displaystyle 32 \over\displaystyle 3}, 4, 
{\displaystyle 2 \over\displaystyle 3})$ 
&  ${\displaystyle 1 \over\displaystyle 3}$ \\ \hline
$(3)$ &
$(XD, ~XD^c)$ &  $({\displaystyle 103 \over\displaystyle 15},
1, -{\displaystyle 7 \over\displaystyle 3})$
 &  ${\displaystyle 25 \over\displaystyle 69}$
& $({\displaystyle 118 \over\displaystyle 15}, 
2, -{\displaystyle 4 \over\displaystyle 3})$
& ${\displaystyle 25 \over\displaystyle 69}$
& $({\displaystyle 148 \over\displaystyle 15},
 4, {\displaystyle 2 \over\displaystyle 3})$ 
&  ${\displaystyle 25 \over\displaystyle 69}$  \\ \hline
$(4)$ &
$(XL, ~XL^c)$ & $(7, {\displaystyle 5 \over\displaystyle 3}, -3)$ 
& ${\displaystyle 7 \over\displaystyle 15}$ 
& $(8, {\displaystyle 8 \over\displaystyle 3}, -2)$ 
& ${\displaystyle 7 \over\displaystyle 15}$ 
&  $(10, {\displaystyle 14 \over\displaystyle 3}, 0)$  
& ${\displaystyle 7 \over\displaystyle 15}$  \\ \hline
$(5)$ &
$(XE, ~XE^c)$ & $({\displaystyle 37 \over\displaystyle 5}, 1, -3)$ 
&  ${\displaystyle 5 \over\displaystyle 13}$

& $({\displaystyle 42 \over\displaystyle 5}, 2, -2)$ 
&  ${\displaystyle 5 \over\displaystyle 13}$
& $({\displaystyle 52 \over\displaystyle 5}, 4, 0)$ 
& ${\displaystyle 5 \over\displaystyle 13}$  \\ \hline
$(6)$ &
$(XY, ~XY^c)$ & $({\displaystyle 149 \over\displaystyle 15}, 3, 
-{\displaystyle 5 \over\displaystyle 3})$ 
& ${\displaystyle 35 \over\displaystyle 87}$
& $({\displaystyle 164 \over\displaystyle 15}, 4, 
-{\displaystyle 2 \over\displaystyle 3})$ 
&  ${\displaystyle 35 \over\displaystyle 87}$
& $({\displaystyle 194 \over\displaystyle 15}, 6, 
{\displaystyle 4 \over\displaystyle 3})$ 
& ${\displaystyle 35 \over\displaystyle 87}$ \\ \hline
$(7)$ &
$XG$ &   $({\displaystyle 33 \over\displaystyle 5}, 1, -1)$ 
& ${\displaystyle 5 \over\displaystyle 19}$ 
 & $({\displaystyle 38 \over\displaystyle 5}, 2, 0)$ 
&  ${\displaystyle 5 \over\displaystyle 19}$ 
& $({\displaystyle 48 \over\displaystyle 5}, 4, 2)$ 
& ${\displaystyle 5 \over\displaystyle 19}$ \\ \hline
$(8)$ &
$XW$ &  $({\displaystyle 33 \over\displaystyle 5},
{\displaystyle 7 \over\displaystyle 3}, -3)$ 
& ${\displaystyle 5 \over\displaystyle 9}$ 
& $({\displaystyle 38 \over\displaystyle 5}, 
{\displaystyle 10 \over\displaystyle 3}, -2)$ 
&  ${\displaystyle 5 \over\displaystyle 9}$
& $({\displaystyle 48 \over\displaystyle 5}, 
{\displaystyle 16 \over\displaystyle 3}, 0)$ 
& ${\displaystyle 5 \over\displaystyle 9}$ \\ \hline
$(9)$ &
$(XT, ~XT^c)$ & $(9, {\displaystyle 11 \over\displaystyle 3}, -3)$ 
& ${\displaystyle 5 \over\displaystyle 9}$ 
& $(10,{\displaystyle 14 \over\displaystyle 3} , -2)$ 
& ${\displaystyle 5 \over\displaystyle 9}$ 
& $(12, {\displaystyle 20 \over\displaystyle 3}, 0)$  
& ${\displaystyle 5 \over\displaystyle 9}$  \\ \hline
$(10)$ &
$(XS, ~XS^c)$ & $({\displaystyle 131 \over\displaystyle 15}, 1, 
{\displaystyle 1 \over\displaystyle 3})$ 
& ${\displaystyle 5 \over\displaystyle 63}$  
& $({\displaystyle 146 \over\displaystyle 15},
 2, {\displaystyle 4 \over\displaystyle 3})$ 
& ${\displaystyle 5 \over\displaystyle 63}$ 
& $({\displaystyle 176 \over\displaystyle 15},
 4, {\displaystyle 10 \over\displaystyle 3})$ 
& ${\displaystyle 5 \over\displaystyle 63}$  \\ \hline
$(11)$ &
$(XQ, ~XQ^c)$   & $({\displaystyle 113 \over\displaystyle 15},
 3, -{\displaystyle 5 \over\displaystyle 3})$ 
& ${\displaystyle 35 \over\displaystyle 69}$ 
& $({\displaystyle 128 \over\displaystyle 15},
 4, -{\displaystyle 2 \over\displaystyle 3})$ 
& ${\displaystyle 35 \over\displaystyle 69}$ 
& $({\displaystyle 158 \over\displaystyle 15},
 6, {\displaystyle 4 \over\displaystyle 3})$ 
& ${\displaystyle 35 \over\displaystyle 69}$  \\ 
 & $(XE, ~XE^c)$ & & & & & & \\ \hline
$(12)$ &
$(XU, ~XU^c)$ & $({\displaystyle 127 \over\displaystyle 15},
 1, -{\displaystyle 7 \over\displaystyle 3})$ 
& ${\displaystyle 25 \over\displaystyle 81}$
& $({\displaystyle 142 \over\displaystyle 15},
 2, -{\displaystyle 4 \over\displaystyle 3})$ 
&  ${\displaystyle 25 \over\displaystyle 81}$
& $({\displaystyle 172 \over\displaystyle 15},
 4, {\displaystyle 2 \over\displaystyle 3})$ 
&   ${\displaystyle 25 \over\displaystyle 81}$ \\ 
 & $(XE, ~XE^c)$ & & & & & & \\ \hline
$(13)$ &
$(XG, ~XW)$ &  $({\displaystyle 33 \over\displaystyle 5},
{\displaystyle 7 \over\displaystyle 3}, -1)$ 
& ${\displaystyle 25 \over\displaystyle 57}$
& $({\displaystyle 38 \over\displaystyle 5},
{\displaystyle 10 \over\displaystyle 3}, 0)$  
& ${\displaystyle 25 \over\displaystyle 57}$
& $({\displaystyle 48 \over\displaystyle 5},
{\displaystyle 16 \over\displaystyle 3}, 2)$ 
& ${\displaystyle 25 \over\displaystyle 57}$ \\ \hline
$(14)$ & $(XT, ~XT^c)$ 
& $({\displaystyle 167 \over\displaystyle 15},
{\displaystyle 11 \over\displaystyle 3},
{\displaystyle 1 \over\displaystyle 3})$ 
& ${\displaystyle 25 \over\displaystyle 81}$
& $({\displaystyle 182 \over\displaystyle 15},
{\displaystyle 14 \over\displaystyle 3},
{\displaystyle 4 \over\displaystyle 3})$ 
&  ${\displaystyle 25 \over\displaystyle 81}$
& $({\displaystyle 212 \over\displaystyle 15},
{\displaystyle 20 \over\displaystyle 3},
{\displaystyle 10 \over\displaystyle 3})$ 
&  ${\displaystyle 25 \over\displaystyle 81}$ \\ 
& $(XS, ~XS^c)$ &  &  & & & & \\ \hline
$(15)$ &
$(\mathbf{5}, \mathbf{\overline{5}})$ 
& $({\displaystyle 109 \over\displaystyle 15},
{\displaystyle 5 \over\displaystyle 3},
-{\displaystyle 7 \over\displaystyle 3})$ 
& ${\displaystyle 5 \over\displaystyle 12}$
& $({\displaystyle 124 \over\displaystyle 15},
{\displaystyle 8 \over\displaystyle 3},
-{\displaystyle 4 \over\displaystyle 3})$ 
& ${\displaystyle 5 \over\displaystyle 12}$
& $({\displaystyle 154 \over\displaystyle 15},
{\displaystyle 14 \over\displaystyle 3},
{\displaystyle 2 \over\displaystyle 3})$ 
& ${\displaystyle 5 \over\displaystyle 12}$ \\ \hline
$(16)$ &
$(\mathbf{10}, \mathbf{\overline{10}})$  
& $({\displaystyle 43 \over\displaystyle 5}, 3, -1)$ 
& ${\displaystyle 5 \over\displaystyle 12}$ 
& $({\displaystyle 48 \over\displaystyle 5}, 4, 0)$  
& ${\displaystyle 5 \over\displaystyle 12}$ 
& $({\displaystyle 58 \over\displaystyle 5}, 6, 2)$  
& ${\displaystyle 5 \over\displaystyle 12}$  \\ \hline
$(17)$ &
$(\mathbf{15}, \mathbf{\overline{15}})$ & 
$({\displaystyle 169 \over\displaystyle 15},
{\displaystyle 17 \over\displaystyle 3},
{\displaystyle 5 \over\displaystyle 3})$ 
& ${\displaystyle 5 \over\displaystyle 12}$
& $({\displaystyle 184 \over\displaystyle 15},
{\displaystyle 20 \over\displaystyle 3},
{\displaystyle 8 \over\displaystyle 3})$ 
& ${\displaystyle 5 \over\displaystyle 12}$
& $({\displaystyle 214 \over\displaystyle 15},
{\displaystyle 26 \over\displaystyle 3},
{\displaystyle 14 \over\displaystyle 3})$ 
& ${\displaystyle 5 \over\displaystyle 12}$   \\ \hline
$(18)$ &
$\mathbf{24}$  &  $({\displaystyle 149 \over\displaystyle 15},
{\displaystyle 13 \over\displaystyle 3},
{\displaystyle 1 \over\displaystyle 3})$  
& ${\displaystyle 5 \over\displaystyle 12}$ 
& $({\displaystyle 164 \over\displaystyle 15},
{\displaystyle 16 \over\displaystyle 3},
{\displaystyle 4 \over\displaystyle 3})$  
& ${\displaystyle 5 \over\displaystyle 12}$ 
& $({\displaystyle 194 \over\displaystyle 15},
{\displaystyle 22 \over\displaystyle 3},
{\displaystyle 10 \over\displaystyle 3})$  
&  ${\displaystyle 5 \over\displaystyle 12}$  \\ \hline
\end{tabular}
\end{center}
\caption{ The parameters $a^0_i$, $a_i^1$, and $a_i^2$, and the 
 indices $k_0$, $k_1$, and $k_2$ of the gaugino mass relations
in the $SU(5)$ models with various  messenger fields. }
\label{KDAM-SUV}
\end{table}










\begin{table}[htb]
\begin{center}
\begin{tabular}{|c|c|c|c|c|||c|c|c|c|c|}
\hline
${\rm Cases }$ &
${\rm Messengers }$ & $k'_0$ & $k'_1$ & $k'_2$ &
${\rm Cases }$ &
${\rm Messengers }$ & $k'_0$ & $k'_1$ & $k'_2$ \\\hline
$(1)$ &
~$(XQ, ~XQ^c)$~ &   ~$121/23$~ & ~$95/39$~ & $14$ &
$(2)$ &
~$(XU, ~XU^c)$~ &  $11/19$ & ~$19/27$~ & $-32/5$ \\ \hline
$(3)$ &
$(XD, ~XD^c)$ &  $11/13$ & $19/21$ & ~$-8/5$~ &
$(4)$ &
$(XL, ~XL^c)$  & $11$ & $19/3$ &   $\infty$ \\ \hline
$(5)$ &
$(XE, ~XE^c)$ & $0$ & $0$ & $\infty$ &
$(6)$ &
$(XY, ~XY^c)$  & ~$121/47$~ & $95/63$ & $22/5$ \\ \hline
$(7)$ &
$XG$ &   $1$ & $1$ & $0$ & 
$(8)$ &
$XW$ &   $\infty$ &   $\infty$ &   $\infty$ \\ \hline
$(9)$ &
$(XT, ~XT^c)$  & $22/3$ &  $38/9$ &  $\infty$ &
$(10)$ &
$(XS, ~XS^c)$  & $55/71$ & $95/111$ & ~$-64/25$~ \\ \hline
$(11)$ &
$(XQ, ~XQ^c)$   & $121/29$ & $19/9$ &  $58/5$ &
$(12)$ &
$(XU, ~XU^c)$ 
& $11/25$ & $19/33$ &  $-56/5$\\ 
 & $(XE, ~XE^c)$  & & & & & $(XE, ~XE^c)$ 
&  & & \\ \hline
$(13)$ &
$XG$ &   $3$ & $5/3$ & $32/5$ &
$(14)$ & $(XT, ~XT^c)$ &  $187/89$ & $57/43$ & $56/25 $ \\ 
 & $XW$ &   &  &  & & 
$(XS, ~XS^c)$ &  &  &  \\ \hline
$(15)$ &
$(\mathbf{5}, \mathbf{\overline{5}})$ &  $11/4$ & $19/12$ & $28/5$ &
$(16)$ &
$(\mathbf{10}, \mathbf{\overline{10}})$  &  $11/4$ & $19/12$ & $28/5$ \\ \hline
$(17)$ &
$(\mathbf{15}, \mathbf{\overline{15}})$ &  $11/4$ & $19/12$ & $28/5$ &
$(18)$ &
$\mathbf{24}$  &  $11/4$ & $19/12$ & $28/5$ \\ \hline
\end{tabular}
\end{center}
\caption{ The  indices $k'_0$, $k'_1$, and $k'_2$ of the gaugino mass relations
in the $SU(5)$ models with various  messenger fields. }
\label{DAM-SUV}
\end{table}



(i) The $SU(5)$ Models \\

We consider three Types of the $SU(5)$ models with or without 
additional SM singlet(s): Type I $SU(5)$ models are the 
minimal $SU(5)$ models ; Type II $SU(5)$ models are the $SU(5)$ models
with TeV-scale vector-like particles 
$(\mathbf{5},~\mathbf{\overline{5}})$;  
Type III $SU(5)$ models are the $SU(5)$ models
with TeV-scale vector-like particles 
$(\mathbf{10},~\mathbf{\overline{10}})$ (or three pairs of
$(\mathbf{5},~\mathbf{\overline{5}})$). We denote the parameters $a_i$,
and the indices $k$ and $k'$ for the gaugino mass
relations 
in Type I $SU(5)$ models as $a_i^0$, $k_0$, and $k'_0$,
in Type II $SU(5)$ models as $a_i^1$, $k_1$, and $k'_1$, and
in Type III $SU(5)$ models as $a_i^2$, $k_2$, and $k'_2$, respectively.
For $k'_i$, we have
\begin{eqnarray}
k'_0 &=& {\displaystyle {33 \sum_j n_2(\Phi_j) g(x_j)
+ 11 \sum_j n_3(\Phi_j) g(x_j)}
\over\displaystyle {5 \sum_j n_1(\Phi_j) g(x_j)
+ 11 \sum_j n_3(\Phi_j) g(x_j) }}  ~,~\, 
\label{SUVkp-0}
\end{eqnarray}
\begin{eqnarray}
k'_1 &=& {\displaystyle {19 \sum_j n_2(\Phi_j) g(x_j)
+ 19 \sum_j n_3(\Phi_j) g(x_j)}
\over\displaystyle {5 \sum_j n_1(\Phi_j) g(x_j)
+ 19 \sum_j n_3(\Phi_j) g(x_j) }}  ~,~\, 
\label{SUVkp-1}
\end{eqnarray}
\begin{eqnarray}
k'_2 &=& {\displaystyle {48 \sum_j n_2(\Phi_j) g(x_j)
-20  \sum_j n_1(\Phi_j) g(x_j)}
\over\displaystyle {
5 \sum_j n_3(\Phi_j) g(x_j) }}  ~.~\, 
\label{SUVkp-2}
\end{eqnarray}

Choosing $p=4$, we present the
parameters $a^0_i$, $a^1_i$, and $a^2_i$,
 and the indices $k_0$,  $k_1$, and $k_2$ 
for various messenger fields
in Table~\ref{KDAM-SUV}. For the Cases $(7)$, $(13)$, and
$(16)$ in Type II $SU(5)$ models, and for the Cases 
$(4)$, $(5)$,  $(8)$, and $(9)$ in Type III $SU(5)$ models,
we have  massless gluino. This problem can be solved
easily by choosing the other $p$, for example, $p=5$.
Also, we present the indices
$k'_0$, $k'_1$, and $k'_2$ for various messenger
fields in  Table~\ref{DAM-SUV}. We emphasize that
the indices $k'_0$, $k'_1$, and $k'_2$ will be the same
if we assume that all the messenger fields have 
the same Yukawa couplings to $X$ since $g(x_j)$ is the same
for all the messenger fields.
 However, in the Cases $(7)$,
$(8)$, and $(13)$, we can not solve the tachyonic slepton problem
since the messenger fields are not charged under $U(1)_Y$. 
Interestingly, the gluino is
the lightest gaugino in most of our scenarios.\\



\begin{table}[htb]
\begin{center}
\begin{tabular}{|c|c|c|c||c|c||c|c|}
\hline
~${\rm Cases }$~ &
~${\rm Messengers }$~ & ~$(a_1^0, ~a_2^0, ~a_3^0)$~  & ~$k_0$~ & 
~$(a_1^1, ~a_2^1, ~a_3^1)$~  & ~$k_1$~ &  
~$(a_1^2, ~a_2^2, ~a_3^2)$~  & ~$k_2$~ \\\hline

$(1)$ &
~$(XF, ~\overline{XF})$~ & $(7, 3, -1)$ 
& ${\displaystyle 1 \over\displaystyle 2}$ 
& $({\displaystyle 38 \over\displaystyle 5}, 6, 2)$
& ${\displaystyle 5 \over\displaystyle 7}$ 
& $({\displaystyle 44 \over\displaystyle 5}, 6, 2)$
& ${\displaystyle 10 \over\displaystyle 17}$  \\ \hline
$(2)$ &
$(Xf, ~\overline{Xf})$ 
& $({\displaystyle 121 \over\displaystyle 15}, 
{\displaystyle 5 \over\displaystyle 3},
- {\displaystyle 7 \over\displaystyle 3})$
& ${\displaystyle 5 \over\displaystyle 13}$ 
&  $({\displaystyle 26 \over\displaystyle 3}, 
{\displaystyle 14 \over\displaystyle 3},
 {\displaystyle 2 \over\displaystyle 3})$
& ${\displaystyle 1 \over\displaystyle 2}$ 
&  $({\displaystyle 148 \over\displaystyle 15}, 
{\displaystyle 14 \over\displaystyle 3},
 {\displaystyle 2 \over\displaystyle 3})$
& ${\displaystyle 10 \over\displaystyle 23}$ \\ \hline
$(3)$ &
$(Xl, ~\overline{Xl})$ 
&  $({\displaystyle 37 \over\displaystyle 5}, 1, -3)$
& ${\displaystyle 5 \over\displaystyle 13}$ 
&  $(8, 4, 0)$ 
&  ${\displaystyle 1 \over\displaystyle 2}$ 
&  $({\displaystyle 46 \over\displaystyle 5}, 4, 0)$ 
&  ${\displaystyle 10 \over\displaystyle 23}$ \\ \hline
$(4)$ &
$(Xh, ~\overline{Xh})$ 
&  $({\displaystyle 109 \over\displaystyle 15}, 
{\displaystyle 5 \over\displaystyle 3},
- {\displaystyle 7 \over\displaystyle 3})$
& ${\displaystyle 5 \over\displaystyle 12}$ 
& $({\displaystyle 118 \over\displaystyle 15}, 
{\displaystyle 14 \over\displaystyle 3},
 {\displaystyle 2 \over\displaystyle 3})$
&  ${\displaystyle 5 \over\displaystyle 9}$
&  $({\displaystyle 136 \over\displaystyle 15}, 
{\displaystyle 14 \over\displaystyle 3},
 {\displaystyle 2 \over\displaystyle 3})$
&  ${\displaystyle 10 \over\displaystyle 21}$ \\ \hline
$(5)$ &
$(XGW, XN)$ 
& $({\displaystyle 101 \over\displaystyle 15}, 
{\displaystyle 13 \over\displaystyle 3},
 {\displaystyle 1 \over\displaystyle 3})$
&  ${\displaystyle 5 \over\displaystyle 8}$
& $({\displaystyle 22 \over\displaystyle 3}, 
{\displaystyle 22 \over\displaystyle 3},
 {\displaystyle 10 \over\displaystyle 3})$
&  $1$
& $({\displaystyle 128 \over\displaystyle 15}, 
{\displaystyle 22 \over\displaystyle 3},
 {\displaystyle 10 \over\displaystyle 3})$
&  ${\displaystyle 10 \over\displaystyle 13}$ \\ \hline
$(6)$ &
$(XX, ~\overline{XX})$ 
&  $({\displaystyle 59 \over\displaystyle 5}, 3, -1)$
&  ${\displaystyle 5 \over\displaystyle 16}$
& $({\displaystyle 62 \over\displaystyle 5}, 6, 2)$
& ${\displaystyle 5 \over\displaystyle 13}$
& $({\displaystyle 68 \over\displaystyle 5}, 6, 2)$
&  ${\displaystyle 10 \over\displaystyle 29}$ \\ \hline
$(7)$ &
$(XF, ~\overline{XF})$  
&   $({\displaystyle 39 \over\displaystyle 5}, 3, -1)$
&  ${\displaystyle 5 \over\displaystyle 11}$
&   $({\displaystyle 42 \over\displaystyle 5}, 6, 2)$
&  ${\displaystyle 5 \over\displaystyle 8}$
&   $({\displaystyle 48 \over\displaystyle 5}, 6, 2)$
&  ${\displaystyle 10 \over\displaystyle 19}$ \\ 
 & $(Xl, ~\overline{Xl})$  &  & &  & & & \\ \hline
$(8)$ &
$(Xf, ~\overline{Xf})$  
& $({\displaystyle 133 \over\displaystyle 15}, 
{\displaystyle 5 \over\displaystyle 3},
- {\displaystyle 7 \over\displaystyle 3})$
&  ${\displaystyle 5 \over\displaystyle 14}$
& $({\displaystyle 142 \over\displaystyle 15}, 
{\displaystyle 14 \over\displaystyle 3},
 {\displaystyle 2 \over\displaystyle 3})$
&  ${\displaystyle 5 \over\displaystyle 11}$
& $({\displaystyle 32 \over\displaystyle 3}, 
{\displaystyle 14 \over\displaystyle 3},
 {\displaystyle 2 \over\displaystyle 3})$
&  ${\displaystyle 2 \over\displaystyle 5}$ \\ 
& $(Xl, ~\overline{Xl})$  &  & & & & & \\ \hline
$(9)$ &
$(Xh, ~\overline{Xh})$ 
& $({\displaystyle 121 \over\displaystyle 15}, 
{\displaystyle 5 \over\displaystyle 3},
- {\displaystyle 7 \over\displaystyle 3})$
&  ${\displaystyle 5 \over\displaystyle 13}$
& $({\displaystyle 26 \over\displaystyle 3}, 
{\displaystyle 14 \over\displaystyle 3},
 {\displaystyle 2 \over\displaystyle 3})$
&  ${\displaystyle 1 \over\displaystyle 2}$
& $({\displaystyle 148 \over\displaystyle 15}, 
{\displaystyle 14 \over\displaystyle 3},
 {\displaystyle 2 \over\displaystyle 3})$
&  ${\displaystyle 10 \over\displaystyle 23}$  \\ 
 & $(Xl, ~\overline{Xl})$ &  &  &  & & & \\ \hline
$(10)$ &
~$(XF, ~\overline{XF})$~ 
& $({\displaystyle 127 \over\displaystyle 15}, 
{\displaystyle 11 \over\displaystyle 3},
- {\displaystyle 1 \over\displaystyle 3})$
&  ${\displaystyle 5 \over\displaystyle 11}$
& $({\displaystyle 136 \over\displaystyle 15}, 
{\displaystyle 20 \over\displaystyle 3},
 {\displaystyle 8 \over\displaystyle 3})$
&  ${\displaystyle 5 \over\displaystyle 8}$
& $({\displaystyle 154 \over\displaystyle 15}, 
{\displaystyle 20 \over\displaystyle 3},
 {\displaystyle 8 \over\displaystyle 3})$
&  ${\displaystyle 10 \over\displaystyle 19}$ \\  
& $(Xf, ~\overline{Xf})$ &  & & & & & \\  \hline
$(11)$ &
$(XF, ~\overline{XF})$ 
& $({\displaystyle 23 \over\displaystyle 3}, 
{\displaystyle 11 \over\displaystyle 3},
- {\displaystyle 1 \over\displaystyle 3})$
&  ${\displaystyle 1 \over\displaystyle 2}$ 
& $({\displaystyle 124 \over\displaystyle 15}, 
{\displaystyle 20 \over\displaystyle 3},
 {\displaystyle 8 \over\displaystyle 3})$
&  ${\displaystyle 5 \over\displaystyle 7}$ 
& $({\displaystyle 142 \over\displaystyle 15}, 
{\displaystyle 20 \over\displaystyle 3},
 {\displaystyle 8 \over\displaystyle 3})$
&  ${\displaystyle 10 \over\displaystyle 17}$ \\
& $(Xh, ~\overline{Xh})$ & & &  & & & \\  \hline
$(12)$ &
$(Xf, ~\overline{Xf})$ 
& $({\displaystyle 131 \over\displaystyle 15}, 
{\displaystyle 7 \over\displaystyle 3},
- {\displaystyle 5 \over\displaystyle 3})$
&  ${\displaystyle 5 \over\displaystyle 13}$ 
& $({\displaystyle 28 \over\displaystyle 3}, 
{\displaystyle 16 \over\displaystyle 3},
 {\displaystyle 4 \over\displaystyle 3})$
&  ${\displaystyle 1 \over\displaystyle 2}$ 
& $({\displaystyle 158 \over\displaystyle 15}, 
{\displaystyle 16 \over\displaystyle 3},
 {\displaystyle 4 \over\displaystyle 3})$
&  ${\displaystyle 10 \over\displaystyle 23}$  \\ 
 & $(Xh, ~\overline{Xh})$ &  &  &  & & & \\  \hline
\end{tabular}
\end{center}
\caption{  The parameters $a^0_i$, $a_i^1$, and $a_i^2$, and the 
 indices $k_0$, $k_1$, and $k_2$ of the gaugino mass relations
in the flipped $SU(5)\times U(1)_X$  models with various  messenger fields.}
\label{KDAM-FSUV}
\end{table}





\begin{table}[htb]
\begin{center}
\begin{tabular}{|c|c|c|c|c||c|c|c|c|c|}
\hline
${\rm Cases }$ &
~${\rm Messengers }$~ & ~$k'_0$~ & ~$k'_1$~ & ~$k'_2$~ &
${\rm Cases }$ &
~${\rm Messengers }$~ &  ~$k'_0$~ & ~$k'_1$~ & ~$k'_2$~
\\\hline
$(1)$ &
~$(XF, ~\overline{XF})$~ &   $11/3$ & $32/5$ &  $38/5$ &
$(2)$ &
$(Xf, ~\overline{Xf})$ &   $2$  & $-8/5$ & $-2/5$ \\ \hline
$(3)$ &
$(Xl, ~\overline{Xl})$ &    $0$ & $\infty$ & $\infty$ &
$(4)$ &
$(Xh, ~\overline{Xh})$ &   $11/4$ & $16/5$ &  $22/5$ \\ \hline
$(5)$ &
$(XGW, XN)$ & ~ $55/14$~ &~ $176/25$~ & ~ $206/25$~ &
$(6)$ &
$(XX, ~\overline{XX})$ &   $11/6$ & ~$-16/5$~ & $-2$  \\ \hline
$(7)$ &
$(XF, ~\overline{XF})$  &  $22/7$ & $24/5$ & $6$ &
$(8)$ &
$(Xf, ~\overline{Xf})$  & $11/7$  & ~$-32/5$~ & ~$-26/5$~ \\ 
 & $(Xl, ~\overline{Xl})$  &  & &  &
 & $(Xl, ~\overline{Xl})$  &  & & \\ \hline
$(9)$ &
$(Xh, ~\overline{Xh})$ &  $2$ & $-8/5$ & $-2/5$ &
$(10)$ &
~$(XF, ~\overline{XF})$~ &  ~$88/29$~ & $22/5$ & $28/5$ \\  
 & $(Xl, ~\overline{Xl})$ &  &  &  &
 & $(Xf, ~\overline{Xf})$ &  &
& \\  \hline
$(11)$ &
$(XF, ~\overline{XF})$ &  $44/13$ & $28/5$ & $34/5$ &
$(12)$ &
$(Xf, ~\overline{Xf})$ &  $44/19$ & $4/5$ & $2$ \\ 
 & $(Xh, ~\overline{Xh})$ & & &  &
 & $(Xh, ~\overline{Xh})$ &  &  & \\  \hline
\end{tabular}
\end{center}
\caption{  The  indices $k'_0$, $k'_1$, and $k'_2$ of the gaugino mass relations
in the flipped $SU(5)\times U(1)_X$ models
with various  messenger fields.}
\label{DAM-FSUV}
\end{table}



(ii) The flipped $SU(5)\times U(1)_X$ models \\

 We  consider three Types of
the flipped $SU(5) \times U(1)_X$ models with 
or without additional SM singlet(s): 
Type I flipped $SU(5) \times U(1)_X$ models are the 
minimal flipped  $SU(5)\times U(1)_X$ models; 
Type II flipped $SU(5) \times U(1)_X$ models are the 
flipped $SU(5) \times U(1)_X$ models
with TeV-scale vector-like particles 
$(XF,~\overline{XF})$;  Type III flipped $SU(5) \times U(1)_X$
 models are the flipped $SU(5) \times U(1)_X$ models
with TeV-scale vector-like particles 
$(XF,~\overline{XF})$ and $(Xl,~\overline{Xl})$. 
Moreover, we denote the parameters $a_i$,
and the indices $k$ and $k'$ for gaugino mass
relations 
in the Type I flipped $SU(5)\times U(1)_X$ models 
as $a_i^0$, $k_0$, and $k'_0$,
in the Type II flipped $SU(5)\times U(1)_X$ 
 models as $a_i^1$, $k_1$, and $k'_1$, and
in the Type III flipped $SU(5)\times U(1)_X$ 
 models as $a_i^2$, $k_2$, and $k'_2$, respectively.
In addition,
$k'_0$ is given by Eq.~(\ref{SUVkp-0}), and we have 
\begin{eqnarray}
k'_1 &=& {\displaystyle {36 \sum_j n_2(\Phi_j) g(x_j)
-20  \sum_j n_1(\Phi_j) g(x_j)}
\over\displaystyle {
5 \sum_j n_3(\Phi_j) g(x_j) }}  ~,~\, 
\label{FSUVkp-1}
\end{eqnarray}
\begin{eqnarray}
k'_2 &=& {\displaystyle {42 \sum_j n_2(\Phi_j) g(x_j)
-20  \sum_j n_1(\Phi_j) g(x_j)}
\over\displaystyle {
5 \sum_j n_3(\Phi_j) g(x_j) }}  ~.~\, 
\label{FSUVkp-2}
\end{eqnarray}

Choosing $p=4$, we present the
parameters $a^0_i$, $a^1_i$, and $a^2_i$,
 and the indices $k_0$,  $k_1$, and $k_2$ 
for various messenger fields
in Table~\ref{KDAM-FSUV}. For the Case $(3)$
in Type II and Type III flipped $SU(5)\times U(1)_X$ models,
we have massless gluino. This problem can be solved
 by choosing the other $p$, for example, $p=5$.
Moreover, we present the indices
$k'_0$, $k'_1$, and $k'_2$ for various messenger
fields in  Table~\ref{DAM-FSUV}.  
We emphasize that
the indices $k'_0$, $k'_1$, and $k'_2$ will be the same
if we assume that all the messenger fields have 
the same Yukawa couplings to $X$.
And we  have gluino as
the lightest gaugino in most of our scenarios.\\



\begin{table}[htb]
\begin{center}
\begin{tabular}{|c|c|c|c||c|c||c|c|}
\hline
~${\rm Cases }$~ &
~${\rm Messengers }$~ & ~$(a_1^0, ~a_2^0, ~a_3^0)$~  & ~$k_0$~ & 
~$(a_1^1, ~a_2^1, ~a_3^1)$~  & ~$k_1$~ &  
~$(a_1^2, ~a_2^2, ~a_3^2)$~  & ~$k_2$~ 
\\\hline
$(1)$ &
$(XFL, ~\overline{XFL})$ 
& $({\displaystyle 107 \over\displaystyle 15}, 
{\displaystyle 11 \over\displaystyle 3}, 
-{\displaystyle 5 \over\displaystyle 3})$ 
& ${\displaystyle 20 \over\displaystyle 33}$ 
& $({\displaystyle 122 \over\displaystyle 15}, 
{\displaystyle 14 \over\displaystyle 3}, 
-{\displaystyle 2 \over\displaystyle 3})$ 
& ${\displaystyle 20 \over\displaystyle 33}$ 
& $({\displaystyle 152 \over\displaystyle 15}, 
{\displaystyle 20 \over\displaystyle 3}, 
{\displaystyle 4 \over\displaystyle 3})$ 
& ${\displaystyle 20 \over\displaystyle 33}$ \\ \hline
$(2)$ &
$(XFR, ~\overline{XFR})$~ 
& $({\displaystyle 131 \over\displaystyle 15}, 
1, -{\displaystyle 5 \over\displaystyle 3})$ 
& ${\displaystyle 10 \over\displaystyle 39}$
& $({\displaystyle 146 \over\displaystyle 15}, 
2, -{\displaystyle 2 \over\displaystyle 3})$ 
& ${\displaystyle 10 \over\displaystyle 39}$
& $({\displaystyle 176 \over\displaystyle 15}, 
4, {\displaystyle 4 \over\displaystyle 3})$ 
& ${\displaystyle 10 \over\displaystyle 39}$ \\ \hline
$(3)$ &
$XD\overline{D}$ 
& $({\displaystyle 103 \over\displaystyle 15}, 
1, -{\displaystyle 7 \over\displaystyle 3})$ 
& ${\displaystyle 25 \over\displaystyle 69}$
& $({\displaystyle 118 \over\displaystyle 15}, 
2, -{\displaystyle 4 \over\displaystyle 3})$ 
& ${\displaystyle 25 \over\displaystyle 69}$
& $({\displaystyle 148 \over\displaystyle 15}, 
4, {\displaystyle 2 \over\displaystyle 3})$ 
& ${\displaystyle 25 \over\displaystyle 69}$  \\ \hline
$(4)$ &
$XL\overline{L}$ 
& $(7, {\displaystyle 5 \over\displaystyle 3}, -3)$ 
& ${\displaystyle 7 \over\displaystyle 15}$
& $(8, {\displaystyle 8 \over\displaystyle 3}, -2)$ 
& ${\displaystyle 7 \over\displaystyle 15}$
& $(10, {\displaystyle 14 \over\displaystyle 3}, 0)$ 
& ${\displaystyle 7 \over\displaystyle 15}$ \\ \hline
$(5)$ &
$(XG4, XWL)$ 
& $({\displaystyle 127 \over\displaystyle 15}, 
{\displaystyle 7 \over\displaystyle 3}, 
-{\displaystyle 1 \over\displaystyle 3})$ 
& ${\displaystyle 10 \over\displaystyle 33}$ 
& $({\displaystyle 142 \over\displaystyle 15}, 
{\displaystyle 10 \over\displaystyle 3}, 
{\displaystyle 2 \over\displaystyle 3})$ 
& ${\displaystyle 10 \over\displaystyle 33}$ 
& $({\displaystyle 172 \over\displaystyle 15}, 
{\displaystyle 16 \over\displaystyle 3}, 
{\displaystyle 8 \over\displaystyle 3})$ 
& ${\displaystyle 10 \over\displaystyle 33}$ \\ 
  & $ XWR$ & &  &  & & &   \\ \hline
$(6)$ &
$XZ$ 
& $({\displaystyle 151 \over\displaystyle 15}, 
5, -{\displaystyle 1 \over\displaystyle 3})$ 
& ${\displaystyle 20 \over\displaystyle 39}$ 
& $({\displaystyle 166 \over\displaystyle 15}, 
6, {\displaystyle 2 \over\displaystyle 3})$ 
& ${\displaystyle 20 \over\displaystyle 39}$ 
& $({\displaystyle 196 \over\displaystyle 15}, 
8, {\displaystyle 8 \over\displaystyle 3})$ 
& ${\displaystyle 20 \over\displaystyle 39}$ \\ \hline
$(7)$ &
$(XFL, ~\overline{XFL})$ 
& $({\displaystyle 37 \over\displaystyle 5}, 
{\displaystyle 11 \over\displaystyle 3}, -1)$ 
& ${\displaystyle 5 \over\displaystyle 9}$ 
& $({\displaystyle 42 \over\displaystyle 5}, 
{\displaystyle 14 \over\displaystyle 3}, 0)$ 
& ${\displaystyle 5 \over\displaystyle 9}$ 
& $({\displaystyle 52 \over\displaystyle 5}, 
{\displaystyle 20 \over\displaystyle 3}, 2)$ 
& ${\displaystyle 5 \over\displaystyle 9}$ \\
 & $XD\overline{D}$  &  &  &  & & & \\ \hline
$(8)$ &
$(XFR, ~\overline{XFR})$~  
& $(9, 1, -1)$ 
& ${\displaystyle 1 \over\displaystyle 5}$
& $(10, 2, 0)$ 
& ${\displaystyle 1 \over\displaystyle 5}$
& $(12, 4, 2)$ 
& ${\displaystyle 1 \over\displaystyle 5}$ \\
& $XD\overline{D}$  &  & & & & & \\ \hline
$(9)$ &
$(XFL, ~\overline{XFL})$ 
& $({\displaystyle 113 \over\displaystyle 15}, 
{\displaystyle 13 \over\displaystyle 3}, 
-{\displaystyle 5 \over\displaystyle 3})$ 
& ${\displaystyle 15 \over\displaystyle 23}$ 
& $({\displaystyle 128 \over\displaystyle 15}, 
{\displaystyle 16 \over\displaystyle 3}, 
-{\displaystyle 2 \over\displaystyle 3})$ 
& ${\displaystyle 15 \over\displaystyle 23}$ 
& $({\displaystyle 158 \over\displaystyle 15}, 
{\displaystyle 22 \over\displaystyle 3}, 
{\displaystyle 4 \over\displaystyle 3})$ 
& ${\displaystyle 15 \over\displaystyle 23}$ \\
 & $XL\overline{L}$   &  &  &  & & & \\ \hline
$(10)$ &
$(XFR, ~\overline{XFR})$~  
& $({\displaystyle 137 \over\displaystyle 15}, 
{\displaystyle 5 \over\displaystyle 3}, 
-{\displaystyle 5 \over\displaystyle 3})$ 
& ${\displaystyle 25 \over\displaystyle 81}$

& $({\displaystyle 152 \over\displaystyle 15}, 
{\displaystyle 8 \over\displaystyle 3}, 
-{\displaystyle 2 \over\displaystyle 3})$ 
& ${\displaystyle 25 \over\displaystyle 81}$
& $({\displaystyle 182 \over\displaystyle 15}, 
{\displaystyle 14 \over\displaystyle 3}, 
{\displaystyle 4 \over\displaystyle 3})$ 
& ${\displaystyle 25 \over\displaystyle 81}$ \\
 & $XL\overline{L}$   &  & & & & & \\ \hline
\end{tabular}
\end{center}
\caption{The parameters $a^0_i$, $a_i^1$, and $a_i^2$, and the 
 indices $k_0$, $k_1$, and $k_2$ of the gaugino mass relations
in the  Pati-Salam $SU(4)_C \times SU(2)_L \times SU(2)_R$
 models with various  messenger fields.}
\label{KDAM-PS}
\end{table}








\begin{table}[htb]
\begin{center}
\begin{tabular}{|c|c|c|c|c||c|c|c|c|c|}
\hline
${\rm Cases }$ &
~${\rm Messengers }$~ & $k'_0$ & $k'_1$ & $k'_2$ &
${\rm Cases }$ &
~${\rm Messengers }$~ & $k'_0$ & $k'_1$ & $k'_2$ 
\\\hline
$(1)$ &
$(XFL, ~\overline{XFL})$ &  $77/13$ & $19/7$ & $88/5$ &
$(2)$ &
$(XFR, ~\overline{XFR})$~ &  $11/19$ &  ~$19/27$ & ~$-32/5$ \\ \hline
$(3)$ &
$XD\overline{D}$ & $11/13$ & $19/21$ & $-8/5$ &
$(4)$ &
$XL\overline{L}$ &  $11$ & $19/3$ & $\infty$ \\ \hline
$(5)$ &
$(XG4, XWL)$ &  $55/29$ & $19/15$ & $2$ &
$(6)$ &
$XZ$ & $121/35$ ~ & ~$95/51$~ & $46/5$ \\ 
 & $ XWR$ & &  &  &
 & &  &  &  \\ \hline
$(7)$ &
$(XFL, ~\overline{XFL})$ &   $55/13$ &  $19/9$ & $56/5$ &
$(8)$ &
$(XFR, ~\overline{XFR})$~  &  $11/17$ &  $19/25$ & $-24/5$ \\ 
 & $XD\overline{D}$  &  &  &  &
 & $XD\overline{D}$  &  & & \\ \hline
$(9)$ &
$(XFL, ~\overline{XFL})$ &  $187/29$~ & $133/45$~ & $106/5$~ &
$(10)$ &
$(XFR, ~\overline{XFR})$~  &  $55/41$ & $1$ & ~$-14/5$~ \\ 
 & $XL\overline{L}$   &  &  &  &
 & $XL\overline{L}$   &  & & \\ \hline
\end{tabular}
\end{center}
\caption{The  indices $k'_0$, $k'_1$, and $k'_2$ of the gaugino mass relations
for various  messenger fields
in the  Pati-Salam $SU(4)_C \times SU(2)_L \times SU(2)_R$ models.}
\label{DAM-PS}
\end{table}


(iii) The Pati-Salam $SU(4)_C\times SU(2)_L\times SU(2)_R$ Models \\

We consider three Types of the Pati-Salam 
$SU(4)_C\times SU(2)_L\times SU(2)_R$ models with or 
without additional SM singlet(s): Type I  Pati-Salam models are the 
minimal  Pati-Salam models; Type II  Pati-Salam models are the Pati-Salam models 
with TeV-scale vector-like particles 
$(\mathbf{5},~\mathbf{\overline{5}})$ under $SU(5)$;
and Type III  Pati-Salam models are the 
 Pati-Salam models
with TeV-scale vector-like particles 
$(\mathbf{10},~\mathbf{\overline{10}})$ (or three pairs of
$(\mathbf{5},~\mathbf{\overline{5}})$) under $SU(5)$.
We denote the parameters $a_i$,
and the indices $k$ and $k'$ for the gaugino mass
relations 
in Type I Pati-Salam models as $a_i^0$, $k_0$, and $k'_0$,
in Type II Pati-Salam models  as $a_i^1$, $k_1$, and $k'_1$, and
in Type III Pati-Salam models as $a_i^2$, $k_2$, and $k'_2$, respectively.
Also, $k'_0$, $k'_1$, and $k'_2$ are given
by Eqs.~(\ref{SUVkp-0}), (\ref{SUVkp-1}), and 
(\ref{SUVkp-2}), respectively.

 Choosing $p=4$, we present the
parameters $a^0_i$, $a^1_i$, and $a^2_i$,
 and the indices $k_0$,  $k_1$, and $k_2$ 
for various messenger fields
in Table~\ref{KDAM-PS}. For the Cases $(7)$ and $(8)$
 in Type II Pati-Salam models, and for the Case 
$(4)$ in Type III Pati-Salam models,
we have  massless gluino. This problem can be solved
 by choosing the other $p$, for example, $p=5$.
Also, we present the indices 
$k'_0$, $k'_1$, and $k'_2$ for various messenger
fields in  Table~\ref{DAM-PS}.   
We emphasize that
the indices $k'_0$, $k'_1$, and $k'_2$ will be the same
if we assume that all the messenger fields have 
the same Yukawa couplings to $X$.
Interestingly, the  gluino is
the lightest gaugino in most of our scenarios.\\

(iv) The Other Possible Models \\

There are some other possible models that are consistent with
gauge coupling unification. For example, in the $SU(5)$ models,
we introduce  one pair of
the vector-like particles $(XD,~XD^c)$ (or $(XL,~XL^c)$)
around the TeV scale, and we introduce two or three or more
pairs of the vector-like particles 
 $(XL,~XL^c)$ (or $(XD,~XD^c)$) at the intermediate scale.
However, to obtain the gauge coupling unification,
we do need to fine-tune the masses of
these vector-like particles. Interestingly, in the
flipped $SU(5)\times U(1)_X$ models, we can relax
 the gauge coupling unification condition due to 
the two-step gauge coupling
unification. Let us present a new kind of the
flipped $SU(5)\times U(1)_X$ models.
We introduce the vector-like particles 
$(Xf,~\overline{Xf})$ around the TeV scale, and introduce
the messenger 
vector-like particles $(XF,~\overline{XF})$ or 
$(XF,~\overline{XF})\oplus (Xh,~\overline{Xh})$ 
at the intermediate scale
 $10^8$ GeV or smaller so that the gauge 
coupling unification can be realized.
For the index $k'$, we have
\begin{eqnarray}
k' &=& {\displaystyle {22 \sum_j n_2(\Phi_j) g(x_j)
+ 22 \sum_j n_3(\Phi_j) g(x_j)}
\over\displaystyle {5 \sum_j n_1(\Phi_j) g(x_j)
+ 22 \sum_j n_3(\Phi_j) g(x_j) }}  ~.~\, 
\end{eqnarray}
For the model with the intermediate-scale 
vector-like messenger fields
 $(XF,~\overline{XF})$, we choose $p=5$.
Thus, we have $a_1=91/10$, $a_2=7/2$, and $a_3=-1/2$,
and the indices $k=5/12$, and $k'=44/23$.
Also, 
for the model with the intermediate-scale 
 vector-like  messenger fields 
 $(XF,~\overline{XF})\oplus (Xh,~\overline{Xh})$, 
we choose $p=4$.
Thus, we have $a_1=148/15$, $a_2=14/3$, and $a_3=2/3$,
and the indices $k=10/23$, and $k'=11/6$.




\section{Implications of Gaugino Mass Relations and Their
Indices }

With the gaugino mass relations and their
indices, we may distinguish the different
supersymmetry breaking mediation mechanisms
and probe the four-dimensional GUTs and
string derived GUTs if we can measure the
gaugino masses at the LHC and future ILC.
In particular, we emphasize again
that the gaugino mass realtions 
in the gravity mediated supersymmetry breaking
is different from those for the gauge and 
anomaly mediated supersymmetry breaking, as discussed
in Section III.
Here, we summarize the indices $k$ of the gaugino mass
relations in the typical GUTs 
with gravity, gauge and anomaly mediated supersymmetry breaking:

\begin{itemize}
{

\item {Gravity Mediated Supersymmetry Breaking}

In the typical four-dimensional $SU(5)$ and $SO(10)$ models,
in the F-theory $SU(5)$ models with $U(1)_Y$ flux, and
in the F-theory $SO(10)$ models with $U(1)_{B-L}$ flux
where the gauge symmetry is broken down to the
$SU(3)_C\times SU(2)_L \times SU(2)_R \times U(1)_{B-L}$
gauge symmetries,
the indices for the gaugino mass relations are either $0/0$ 
 or $5/3$, where $k=0/0$ means the mSUGRA gaugino mass 
relation~\cite{Li:2010xr}.
However,  in the F-theory $SO(10)$ models with $U(1)_{X}$ flux
where the gauge symmetry is broken down to the flipped
$SU(5)\times U(1)_X$ gauge symmetries,
we only have the mSUGRA gaugino mass relation~\cite{Li:2010xr}.
Also, in the four-dimensional minimal $SO(10)$ model~\cite{Barr:1997hq}, 
the Higgs field,
which breaks the $SO(10)$ gauge symmetry, is in the $SO(10)$
 $\mathbf{45}$ representation.  Thus, only the dimension-six operators
can induce the non-universal SM gauge kinetic functions
 at the GUT scale, and then such non-universal effects on the SM 
gauge kinetic functions are very small and negligible. Therefore,
we only have the mSUGRA gaugino mass relation as well.
In short, if we obtain $k=5/3$ from the LHC and ILC experiments,
we can rule out the  F-theory $SO(10)$ models with $U(1)_{X}$ flux
and the four-dimensional minimal $SO(10)$ model.

\item {Gauge Mediated Supersymmetry Breaking }

In the four-dimensional $SU(5)$ and $SO(10)$ models, we have
the mSUGRA gaugino mass relation in general since it is difficult
to split the complete $SU(5)$ and $SO(10)$ multiplets. However,
in the orbifold GUTs and F-theory GUTs with various
messenger fields, we have many new
possible gaugino mass relations and their indices, as discussed in Section IV.
In particular, the indices $k$ can be $5/3$ in quite a few 
$SU(5)$ models and Pati-Salam models.
In the flipped $SU(5)\times U(1)_X$ models, we have $k=0$
in general, which are different from the mSGURA gaugino mass
relation except that the messenger fields are $Xh$ and 
$\overline{Xh}$.

\item { UV Insensitive Anomaly Mediated Supersymmetry Breaking}

In the four-dimensional $SU(5)$ and $SO(10)$ models (or say
Pati-Salam models) with
or without the TeV-scale vector-like particles
that form complete GUT multiplets, we generically
have $k=5/12$. In the flipped $SU(5)\times U(1)_X$ models,
in addition to $k=5/12$, we can have $k=5/9$,  $k=10/27$, 
and $k=10/21$. Especially, all the indices $k$ are smaller than 1,
and then they can not be $5/3$ as in the gravity mediated
supersymmetry breaking.

\item {Deflected Anomaly Mediated Supersymmetry Breaking }

If the messenger fields form complete $SU(5)$ or $SO(10)$
representations, we also have $k=5/12$. For generical
messenger fields, the detailed discussions are given
in subsection V.B. Especially, all the indices $k$ are smaller 
than 1. In addition, we would like to point 
out that the discussions for mirage mediation~\cite{mirage}
are similar to those for the deflected AMSB.

}
\end{itemize} 

Furthermore, to distinguish the different scenarios with the
same gaugino mass relations and the same indices,
we need to consider the squark and slepton
masses as well, which will be studied elsewhere~\cite{TLDN}.




\section{Conclusions}

In  GUTs from  orbifold constructions, 
intersecting D-brane model building on Type II orientifolds,
M-theory on $S^1/Z_2$ with Calabi-Yau compactifications, and 
F-theory with $U(1)$ fluxes, we pointed out that
 the generic vector-like
particles do not need to form the complete $SU(5)$ or $SO(10)$ 
representations. Thus, in the GMSB and deflected AMSB,
the messenger fields do not need to form complete
 $SU(5)$ representations. We can achieve the gauge coupling
unification by introducing  the extra vector-like particles that
do not mediate supersymmetry breaking. To be concrete,
we presented the  orbifold $ SU(5)$ models with 
additional vector-like particles, the  orbifold $SO(10)$ models with 
additional vector-like particles where the gauge symmetry
can be broken down to  flipped $SU(5)\times U(1)_X$ or
 Pati-Salam $SU(4)_C \times SU(2)_L \times SU(2)_R$
gauge symmetries, and the  F-theory $SU(5)$ models
with generic vector-like particles. 
Interestingly, these vector-like particles can be the TeV-scale 
vector-like particles that we need to increase the 
lightest CP-even Higgs boson mass in the MSSM, and they can be
the messenger fields in the GMSB and deflected AMSB as well.

In addition, we have studied the general gaugino mass relations
and their indices in the GMSB and AMSB, which are
valid from the GUT scale to the electroweak scale at one loop.
For the GMSB, we  calculated the gaugino
mass relations and their indices for the $SU(5)$ models, the 
flipped $SU(5)\times U(1)_X$ models, and the  Pati-Salam 
$SU(4)_C \times SU(2)_L \times SU(2)_R$ models with various possible
messenger fields. These kinds of GUTs can be realized in 
orbifold GUTs, F-theory $SU(5)$ models with $U(1)_Y$ flux,
and F-theory $SO(10)$ models with $U(1)_X$ flux and $U(1)_{B-L}$
flux where the $SO(10)$ gauge symmetry is respectively broken down
to the flipped $SU(5)\times U(1)_X$ gauge symmetries
and the $SU(3)_C\times SU(2)_L \times SU(2)_R \times U(1)_{B-L}$
gauge symmetries.  Especially, we pointed out that
using gaugino mass relations and their indices, 
we may probe the messenger fields at the intermediate scale.
Moreover, for the AMSB, we considered 
the UV insensitive AMSB and the deflected AMSB. 
In the UV insensitive AMSB, we calculated the gaugino mass
relations and their indices in the $SU(5)$ models without and
with TeV-scale vector-like particles that form complete
$SU(5)$ multiplets, and in the flipped $SU(5)\times U(1)_X$
models with  TeV-scale vector-like particles that form complete
$SU(5)\times U(1)_X$ multiplets.  To achieve the one-step gauge coupling
unification, we emphasize that the discussions for the
Pati-Salam models are similar to those in the $SU(5)$ models.
 In the deflected AMSB,  we defined
the new indices for the gaugino mass relations to 
probe the messenger fields at intermediate scale.
Without or with
the suitable TeV-scale vector-like particles that can lift
the lightest CP-even Higgs boson mass,
we studied the generic gaugino mass relations, and their
 indices $k$ and $k'$ in  the $SU(5)$ models, the 
flipped $SU(5)\times U(1)_X$ models, and the  Pati-Salam 
$SU(4)_C \times SU(2)_L \times SU(2)_R$ models with various possible
messenger fields. Also, we found that in most of our scenarios,
gluino can be the lightest gaugino at low energy.
Especially, we proposed a new kind of interesting flipped
$SU(5)\times U(1)_X$ models. 

Furthermore, using the gaugino mass relations 
and their indices, we may not only determine 
the  supersymmetry breaking mediation mechanisms,
  but also probe the four-dimensional
GUTs, orbifold GUTs, and F-theory GUTs.

\begin{acknowledgments}


This research was supported in part 
by the Natural Science Foundation of China 
under grant numbers 10821504 and 11075194 (TL),
and by the DOE grant DE-FG03-95-Er-40917 (TL and DVN).

\end{acknowledgments}




\appendix

\section{Breifly Review of del Pezzo Surfaces}

The del Pezzo surfaces $dP_n$, where $n=1,~2,~...,~8$, are
defined by blowing up $n$ generic points of 
$\mathbb{P}^{1}\times\mathbb{P}^{1}$ or 
$\mathbb{P}^2$. The homological group
$H_2(dP_n, Z)$ has the generators
\begin{equation}
H,~E_1, ~E_2,~...,~E_n~,~\,
\end{equation}
where $H$ is the hyperplane class for $P^2$, and $E_i$ are the
exceptional divisors at the blowing up points and are
 isomorphic to $\mathbb{P}^{1}$.
 The intersecting numbers of the generators are
\begin{equation}
H\cdot H=1~,\;\;\:E_{i}\cdot E_{j}=-\delta_{ij}~,\;\;\;H\cdot E_{i}=0~.~\,
\end{equation}
The canonical bundle on $dP_{n}$ is given by
\begin{equation}
K_{dP_{n}}=-c_{1}(dP_{n})=-3H+\sum_{i=1}^{n}E_{i}.
\end{equation}
For $n\geq3$,  we can define the generators as follows
\begin{equation}
\alpha_i=E_i-E_{i+1}~,~~~{\rm where}~~i=1,~2,...,~n-1~,~\,
\end{equation}
\begin{equation}
\alpha_n=H-E_1-E_2-E_3~.~\,
\end{equation}
Thus, all the generators $\alpha_i$ is perpendicular to the canonical
class $K_{dP_{n}}$. And
 the intersection products are equal to the negative Cartan
matrix of the Lie algebra $E_n$, and can be considered as simple 
roots. 

The  curves $\Sigma_i$ in $dP_n$ where the particles are localized 
 must be divisors of $S$. And the genus for curve $\Sigma_i$ is
given by 
\begin{equation}
2 g_i-2~=~[\Sigma_i]\cdot ([\Sigma_i]+K_{dP_{k}})~.~\,
\end{equation}

For a line bundle $L$ on the surface $dP_{n}$ with
\begin{equation}
c_{1}(L)=\sum_{i=1}^{n}a_{i}E_{i},
\end{equation}
where $a_{i}a_{j}<0$ for some $i\neq j $,  the K\"ahler form
$J_{dP_{n}}$ can be constructed as follows
\cite{Beasley:2008dc}
\begin{equation}
J_{dP_{k}}=b_0H-\sum_{i=1}^{n}b_{i}E_{i},
\end{equation}
where $\sum_{i=1}^k a_{i}b_{i}=0$ and $b_0 \gg b_{i}>0$. By the
construction, it is easy to see that the line bundle $L$ solves
the BPS equation $J_{dP_k}\wedge c_{1}(L)=0$.




\end{document}